\documentclass[12pt,preprint]{aastex}

\shorttitle{Late-Type Red Supergiants}
\shortauthors{Levesque et al}
\begin{document}

\title{Late-Type Red Supergiants: Too Cool for the Magellanic Clouds?}

\author{Emily M. Levesque\altaffilmark{1}}
\affil{Institute for Astronomy, University of Hawaii, 2680 Woodlawn Drive,
Honolulu, HI 96822}
\email{emsque@ifa.hawaii.edu}

\author {Philip Massey\altaffilmark{1}}
\affil{Lowell Observatory, 1400 W. Mars Hill Road, Flagstaff, AZ 86001}
\email{Phil.Massey@lowell.edu}

\author{K. A. G. Olsen}
\affil{Cerro Tololo Inter-American Observatory, National Optical Astronomy Observatory, 
Casilla 603, La Serena, Chile}
\email{kolsen@noao.edu}

\author{Bertrand Plez}
\affil{GRAAL, Universite Montpellier II, CNRS, 34095 Montpellier, France}
\email{Bertrand.Plez@graal.univ-montp2.fr}

\altaffiltext{1}{Visiting Astronomer, Cerro Tololo Inter-American Observatory (CTIO), 
National Optical Astronomy Observatory (NOAO), which is operated by the Association of 
Universities for Research in Astronomy (AURA), Inc., under cooperative agreement with the 
National Science Foundation (NSF).}

\begin{abstract}

We have identified seven red supergiants (RSGs) in the Large Magellanic Cloud (LMC) and four
RSGs in the Small Magellanic Cloud (SMC), all of which have spectral types that are considerably
later than the average type observed in their parent galaxy.
Using moderate-resolution optical spectrophotometry and the MARCS stellar
atmosphere models, we determine their physical properties and place them on the H-R diagram
for comparison with the predictions of current stellar evolutionary tracks. 
The radial velocities of these stars suggest that they are likely
all members of the Clouds rather than foreground dwarfs or halo giants. Their
locations in the H-R diagram also show us that these stars are cooler than the current evolutionary
tracks allow, appearing to the right of the Hayashi limit, a region in which
stars are no longer in hydrodynamic equilibrium. These stars exhibit considerable variability in their $V$ magnitudes, and three of these
stars also show changes in their effective temperatures (and spectral
types) on the time-scales of months. One of these stars, [M2002] SMC 055188, was caught in
an M4.5 I state, as late as that seen in HV 11423 at its recent extreme:
considerably later, and cooler, than any other supergiant in the SMC. In
addition, we find evidence of variable extinction due to circumstellar dust
and changes in the stars' luminosities,
also consistent with our recent findings for HV 11423 - when
these stars are hotter they are also dustier and more luminous.
 We suggest that these
stars have unusual properties because they are in an unstable (and short-lived)
evolutionary phase.
\end{abstract}


\section{Introduction}
\label{Sec-intro}

Red supergiants (RSGs) are a He-burning phase in the evolution of moderately high mass
stars (10-25$M_\odot$).
Until recently, the location of RSGs on the Hertzsprung-Russell diagram was at odds with predictions
of stellar evolutionary tracks. Levesque et al.\ (2005, hereafter Paper I) fitted the new generation
of MARCS atmosphere models (Gustafsson et al.\ 1975, 2003; Plez et al.\ 1992; Plez 2003) to
moderate-resolution optical spectrophotometry of Galactic RSGs. The physical parameters derived
from this work brought the stars into much better agreement with the Geneva evolutionary tracks
for solar metallicity (Meynet \& Maeder 2003).  Subsequently we performed a similar analysis
(Levesque et al.\ 2006, hereafter Paper II) 
for RSGs in the Magellanic Clouds, where the metallicity is significantly
lower ($Z/Z_\odot=0.5$ for the LMC, and $Z/Z_\odot$=0.2 for the SMC; see Westerlund 1997), with
a similar improvement seen.

Paper~II emphasized that, on average, 
RSGs in the Magellanic Clouds were not as cool as Galactic RSGs,
in accordance with the shifting of the right-most extension (the Hayashi limit, as described in Hayashi \& Hoshi 1961) of the evolutionary tracks 
to warmer effective temperatures at lower metallicities.  Stars in this region are fully convective,
and cooler stars would not be in hydrostatic equilibrium.   This fact is reflected in the
shifting of the observed average spectral subtypes of RSGs in these
galaxies, from M2~I  in the Milky Way to M1~I in the LMC and
K5-7~I in the SMC (Massey \& Olsen 2003), in accordance with the observation and explanation originally
put forth by Elias et al.\ (1985).

However, Massey \& Olsen (2003) did find a few SMC and LMC RSGs that were
considerably later in type than average.   Radial velocities (Massey \& Olsen 2003) of these
spectral outliers
had suggested these could not be foreground Galactic dwarfs, but could
not rule out the possibility that they were Galactic halo giants, as their
radial velocities would be similar to confirmed SMC and LMC 
members.  Were these Magellanic Cloud members?  Or did they present a challenge to evolutionary
theory?  We identified additional late-type RSG candidates from our
own previously unpublished spectroscopy (from late 2004, described below) and/or
broad-band colors ($V-K$, primarily) and decided to re-investigate their membership 
and physical properties.

Here we present moderate-resolution spectrophotometry of seven late-type RSGs in the LMC
and four late-type RSGs in the SMC (\S~\ref{Sec-obs}).  In analyzing the
data (\S~\ref{Sec-analysis}), we begin by first considering the question of whether
or not these stars are members
in the Clouds (\S~\ref{Sec-RVs}). Next, we determine their spectral types and 
physical properties
(\S\ref{Sec-spec}) from our spectrophotometry.
 As a ``reality check",   we also derive the physical properties from broad-band
$V-K$ photometry (\S\ref{Sec-bb}). In order to better understand the nature of
these objects, we also examine their photometric and spectral
variability (\S\ref{Sec-var}).
In \S\ref{Sec-results} we place these stars on the HR diagram, and also revise our 
metallicity-dependent effective
temperature calibration of RSGs based on the new data.  In \S\ref{Sec-conclude} we summarize our findings and discuss our future work on the subject.

\section{Observations and Analysis}
\subsection{Observations}
\label{Sec-obs}

We list our sample of late-type Magellanic Cloud RSGs in Table~\ref{tab:stars}.  
Observations were made on 20-23 December 2005 using the RC  
Spectrograph on the 4-m 
Blanco telescope. The detector was a Loral CCD (3000 x 1000 pixel).
We used a 316 line mm$^{-1}$ 
grating (KPGL2) in first order, which gave  us 2\AA\ pixel$^{-1}$.  A 225 $\mu$m 
(1.5") slit was used, yielding a  spectral resolution of 7.5\AA\ (3.8 pixels).  We 
observed with two  different grating tilts, one covering 3400-6200\AA, and one 
covering 5300-9000\AA.  For the  blue observations we used a BG-39
blocking filter to eliminate any possibility of scattered red light  affecting our 
observations, particularly in the far blue and near-UV, where the stellar flux is 
small; in the red, we  used a WG-495 filter to block out second-order blue light. 
Exposure times ranged from 200-900 seconds in the blue, and  150-600 seconds in the red. 

 Observations were made at the parallactic angle, to ensure good flux calibration, 
and numerous spectrophotometric standard stars (chosen from the 
list of Hamuy et al.\ 1992) were also observed.  We observed several ``featureless"
stars in order to remove telluric absorption,  following Bessell (1990). The reduced blue and red
spectra were combined in order to bring the flux levels into agreement.

For six of our stars (SMC 046662, SMC 055188, LMC 143035, LMC 150020, LMC 158646, and LMC 170452)
we also had full or partial spectrophotometry obtained during November and December 2004 with the
same instrument. (The details of that instrumental setup are given in Paper II.) For SMC 046662
and SMC 055188 the data are complete from 4100\AA\ to 9100\AA. For the other four stars
 observed in 2004, we have
only partial data in the ``blue'' (4100-6450\AA). Although we could not use the latter for
determining physical properties (as the baseline was insufficient for accurate extinction
determinations), these data nevertheless proved very useful, as it provided a means for
checking for spectral variability and determining the relevant timescales. For one of our stars
(SMC 083593) we had incomplete data during both observing runs: the blue was observed in 2004,
and the red was obtained during this paper's primary observing run in 2005. The slopes and line depths
agreed well in the region of overlap, and so we combined the two halves for our analysis of the spectra.

\section{Analysis}
\label{Sec-analysis}
We classify all of the stars in this sample (from the 2005 data and, when available, the 2004 data), based
primarily on the TiO band depths. While all
of the TiO bands are examined, the 
TiO bands at $\lambda \lambda$ 5847, 6158, and 6658 all become
distinctively and progressively stronger at later spectral types, and
as a result are the primary features considered in our classification. For most of the 2004 observations
our data extend only up to 6450\AA, and so we made do with the first two bands.
Our method of classification was consistent with that used in Papers I and II.

\subsection{Membership in the Clouds}
\label{Sec-RVs}
A magnitude- and color-selected sample of RSG candidates will be contaminated by  foreground Galactic 
dwarfs and potentially by more distant halo dwarfs and giants.  
Of these, only the foreground dwarfs can be easily recognized on the basis
of radial velocities, but these are by far the major contaminant in studying RSGs
in Local Group galaxies (see, for instance, Massey 1998). Massey \& 
Olsen (2003) obtained precision radial  velocities, with the CTIO 4-m and Hydra 
fiber positioner feeding  a bench-mounted
spectrograph, of a sample of red stars seen towards the Clouds.
Indeed most of the stars in their sample had radial velocities consistent with those
of the Magellanic Clouds, plus a small fraction (11\% for the SMC, and 5.3\% for the LMC) which
had much smaller radial velocities.  The latter are readily identified as foreground dwarfs,
while the former are tentatively identified as 
{\it bona fide} RSGs.   However, since most
of the SMC's and LMC's apparent radial velocity is simply a reflection of
the sun's motion, the sample of RSGs could be contaminated by red stars
in the Milky Way's halo.  In Paper~II we estimated that this would be a few percent or
less, but here we reconsider the issue.

The sample of stars in Paper~II and here mostly have $12<V<14$, with a few fainter stars.
Their $B-V$ colors are greater than 1.6 (see Table 1 of Paper~II and Table 1 of the present
paper).  According to an updated version of the Bahcall \& Soniera (1980) model, kindly
provided by Heather Morrison, we expect a surface density of  halo stars (all giants)
of about $0.2\pm0.15$ deg$^{-2}$ in this magnitude/color range towards either the LMC 
or the SMC, where the uncertainty reflects
the effects of different assumptions.  The area of the Massey (2002) survey was 14.5 deg$^2$
towards the LMC, and 7.2 deg$^2$ towards the SMC, so we might expect $2.9\pm2.1$ halo
giants (0.6\% of the bright and red stars) seen towards the LMC, and $1.4\pm1.1$ (0.9\%)
seen towards the SMC in his catalog.   Thus we expect only a fraction of a star in the entire
spectroscopic sample of 84 stars (73 discussed in Paper~II and 11 discussed here). 
Of the 11 stars in Table~1, 1\% (0.1 stars) is probably a large overestimate, as the vast majority
of these few halo contaminates have $B-V<1.8$, while all but two of the stars in Table~1 have
$B-V>1.8$.

\begin{figure}
\epsscale{1.0}
\plotone{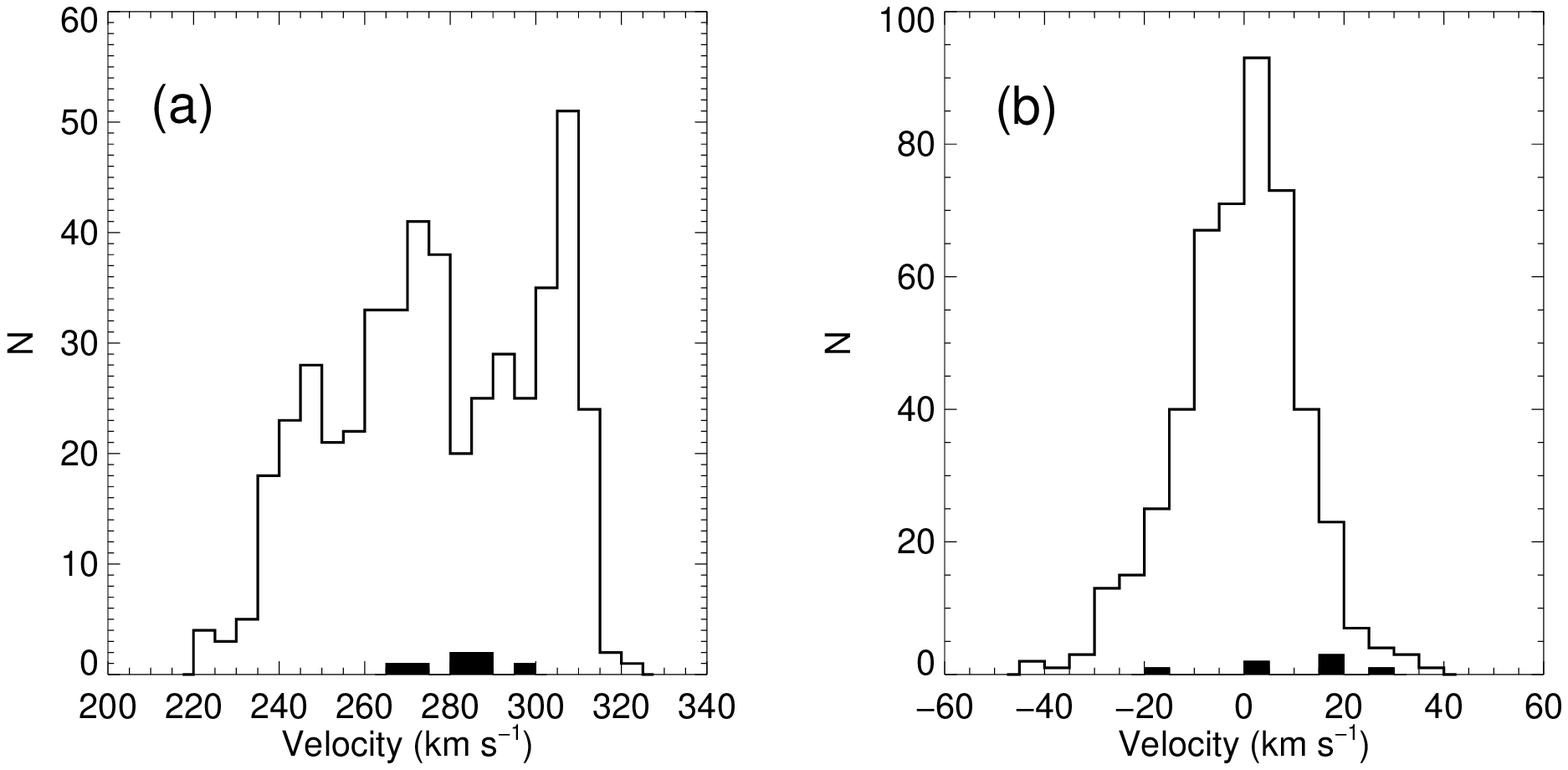}
\caption{\label{fig:kinematics}Kinematics of the LMC. Figure (a) shows a histogram of LMC RSG velocities analyzed in Olsen \& Massey (2007), in bins of 5 km s$^{-1}$, with our seven LMC stars overplotted as a solid histogram, showing that the velocities are consistent with those of other RSGs in the LMC. Figure (b) shows the velocity residuals of our RSGs after subtracting a kinematic model fit to the Olsen \& Massey (2007) RSG sample, with our LMC stars again plotted as a solid histogram. Standard deviation of the velocity residuals is 12 km s$^{-1}$ for the larger sample and 14 km s$^{-1}$ for our stars.}
\end{figure}

Finally, we can use a more exacting test of the kinematics than the simple 
radial velocity cutoff used by Massey \& Olsen (2003), and
ask if the LMC stars discussed here follow the radial velocities of other RSGs as a function of
spatial position in the LMC.  
The kinematics of the SMC are quite complex, and so we restrict the argument here to the LMC, where the kinematics
are relatively well understood (Olsen \& Massey 2007).  Recall that differences in the radial velocities
within the LMC are dominated by the transverse motion of the LMC coupled to
the change in position factor due to the LMC's large angular extent, and that other effects,
such as rotation, are a relatively minor perturbation (see, for example, Meatheringham et al. 1988, Schommer et al. 1992, and van der Marel et al. 2002).  Fig.~\ref{fig:kinematics} shows a histogram of LMC RSG velocities analyzed in Olsen \& Massey (2007). The LMC RSGs discussed here follow the kinematics
of the galaxy, something we do not expect to be true of halo giants.  We conclude that the sample we discuss here
is unlikely to contain foreground objects.

\subsection{Modeling the Spectrophotometry}
\label{Sec-spec}

The observed spectral energy distributions of the sample stars were compared to MARCS
stellar atmosphere models of metallicity $Z/Z_{\odot}=0.2$ for the SMC and $Z/Z_{\odot}=0.5$ for the LMC.
The models ranged from 3000 to 4500 K in increments of 100 K, and were interpolated for
intermediate temperatures at 25 K increments. The log $g$ values for the models ranged from
$-1$ to $+1$ in increments of 0.5 dex.

When fitting the data, we reddened the models using a Cardelli et al.\ (1989) reddening
law with $R_V = 3.1$. The reddening and $T_{\rm eff}$ for each object was determined
by finding the best by-eye fit to both the spectral features and continuum, initially using
a model with log $g = 0.0$. For these later-type
stars with distinct TiO features, our precision was about 50 K, while for the earlier
K-type stars from Paper I, our precision was approximately 100 K. The extinction values $A_V$ are
determined to 0.15 mag. Upon fitting the reddening and $T_{\rm eff}$, we then examined the appropriate
log $g$ value for the star: the bolometric corrections from the models were used with the reddening
and photometry to compute the bolometric luminosity, assuming distance moduli of 18.9 for the SMC
and 18.5 for the LMC (van den Bergh 2000). From the bolometric luminosity and $T_{\rm eff}$, the physical log g was determined.
If the resulting value indicated that a model with log $g = -0.5$ or $+0.5$ would be more
appopriate than the initial estimate of log $g = 0.0$, the star was refitted with the more
appropriate surface gravity value and the process was repeated. A changed log $g$ value did not
affect the $T_{\rm eff}$ determination, but did slightly affect the extinction value. Typically this
process converged upon a satisfactory log $g$ choice for the model after two or three iterations.  

\begin{figure}
\epsscale{0.3}
\plotone{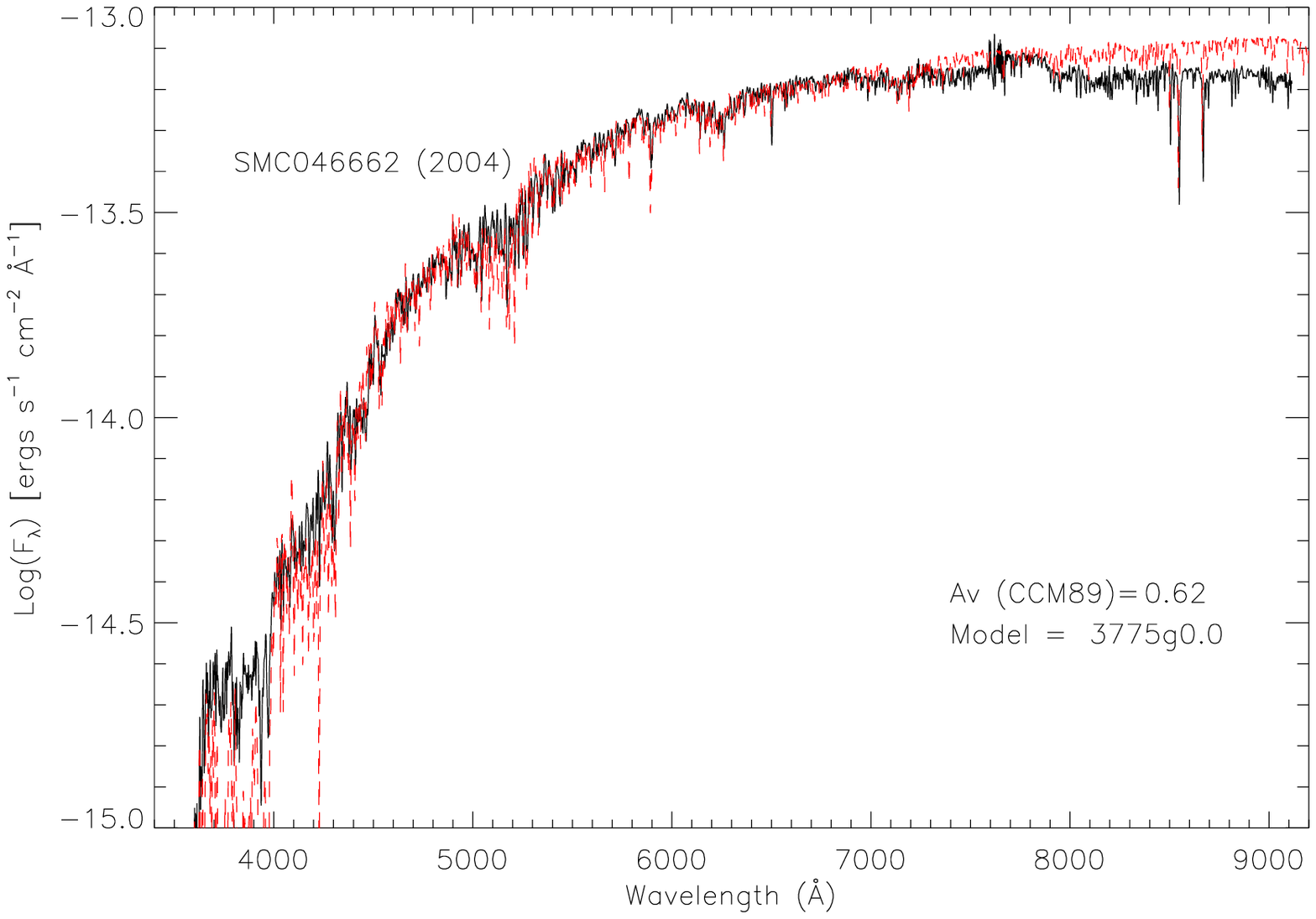}
\plotone{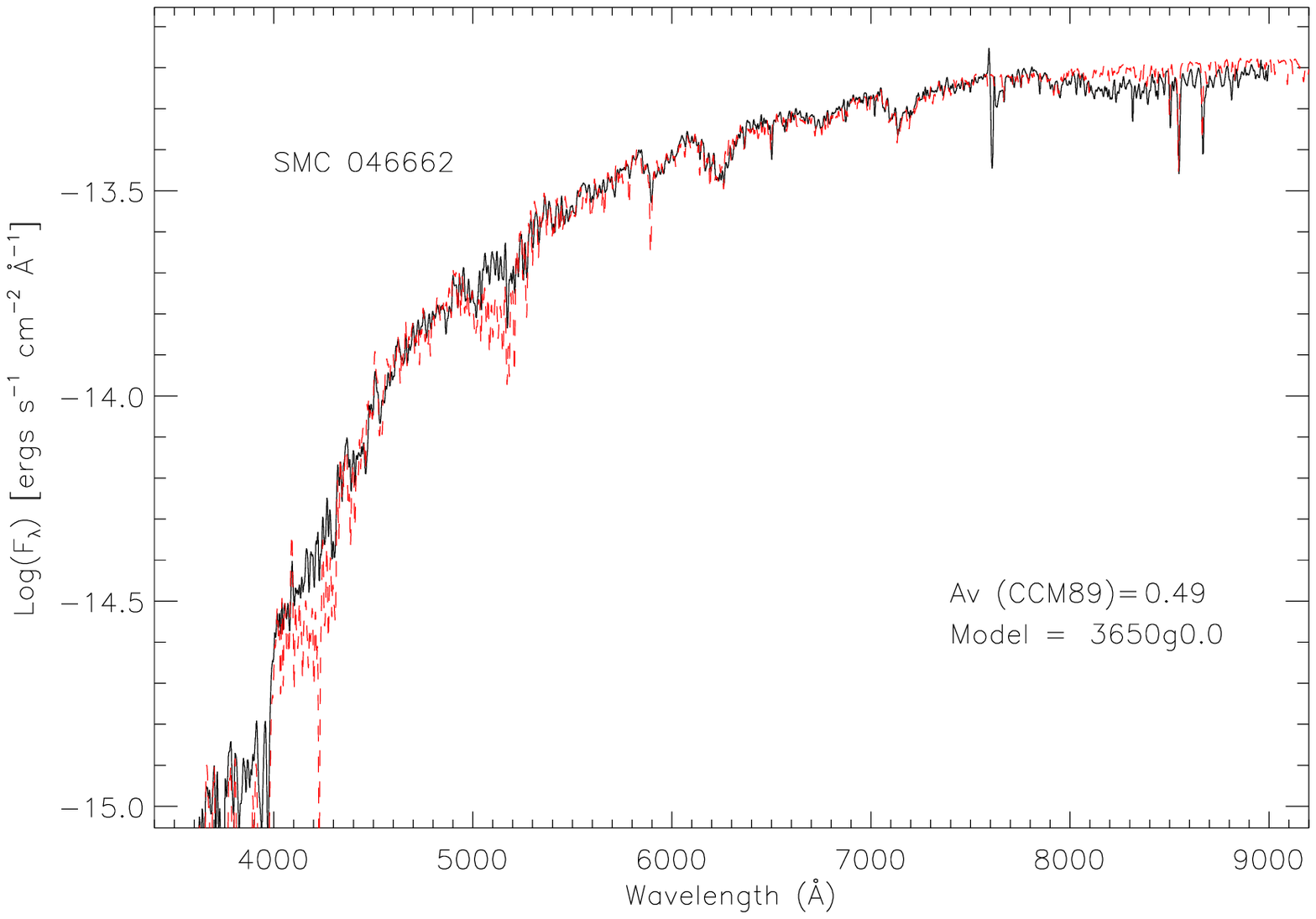}
\plotone{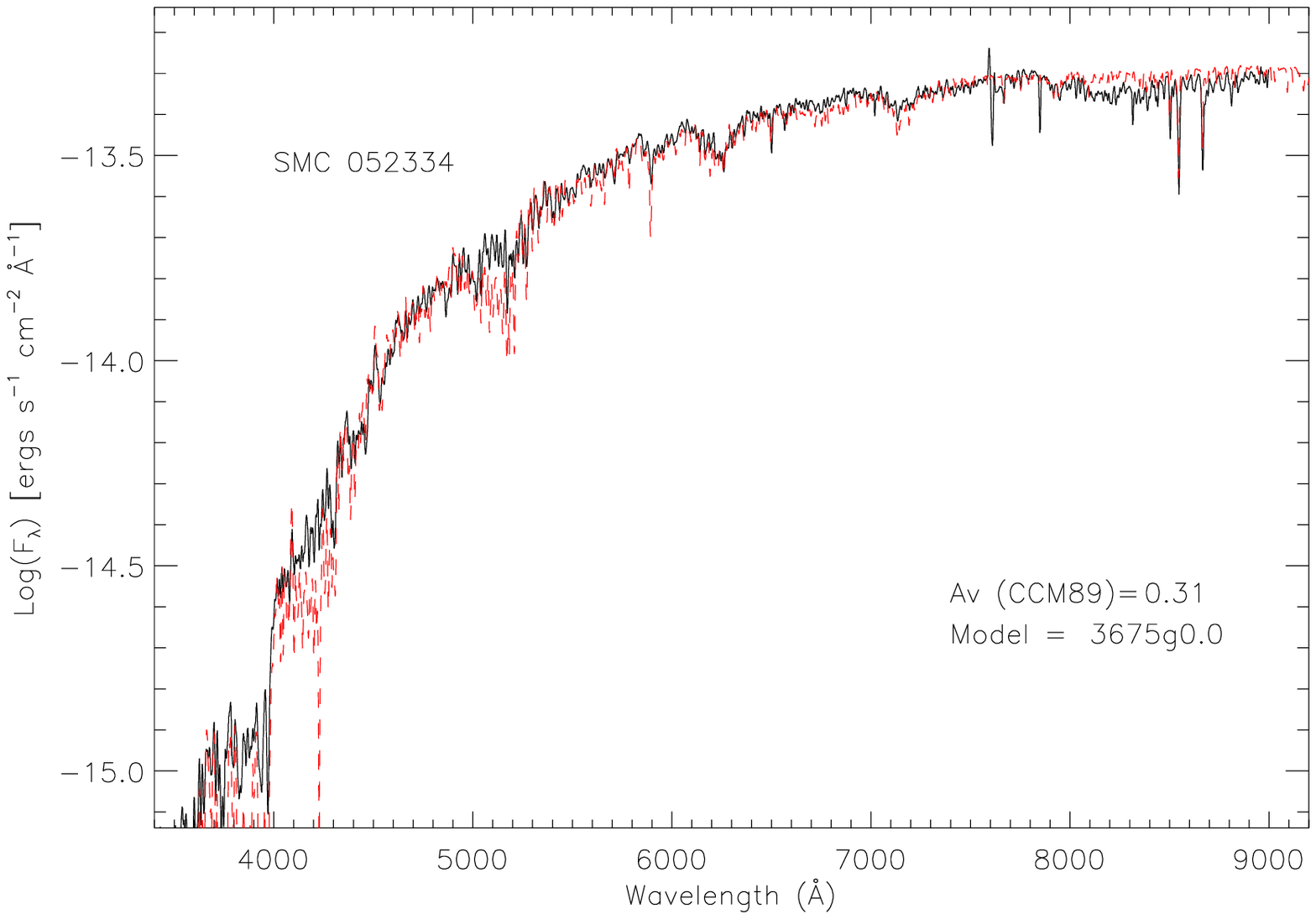}
\plotone{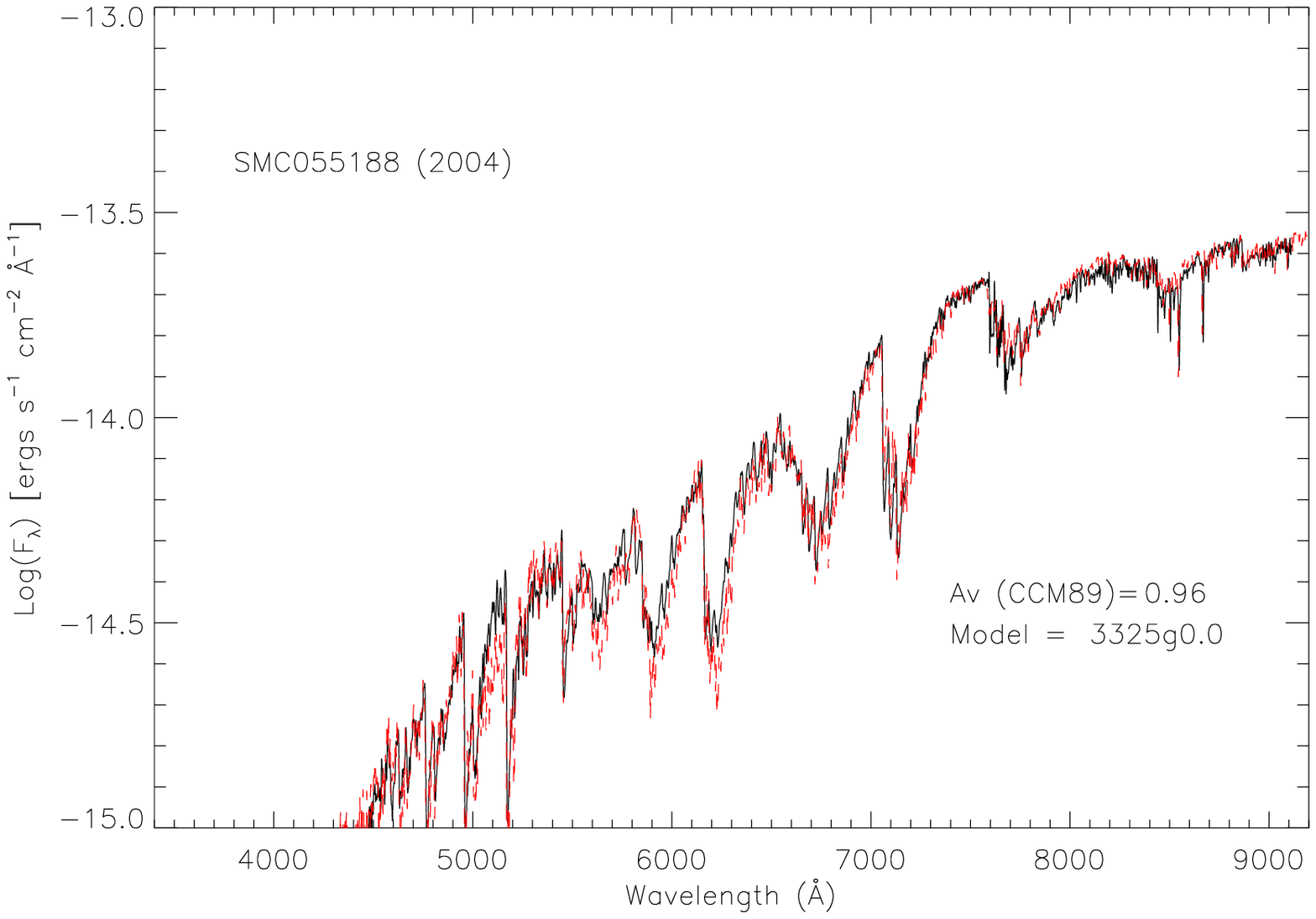}
\plotone{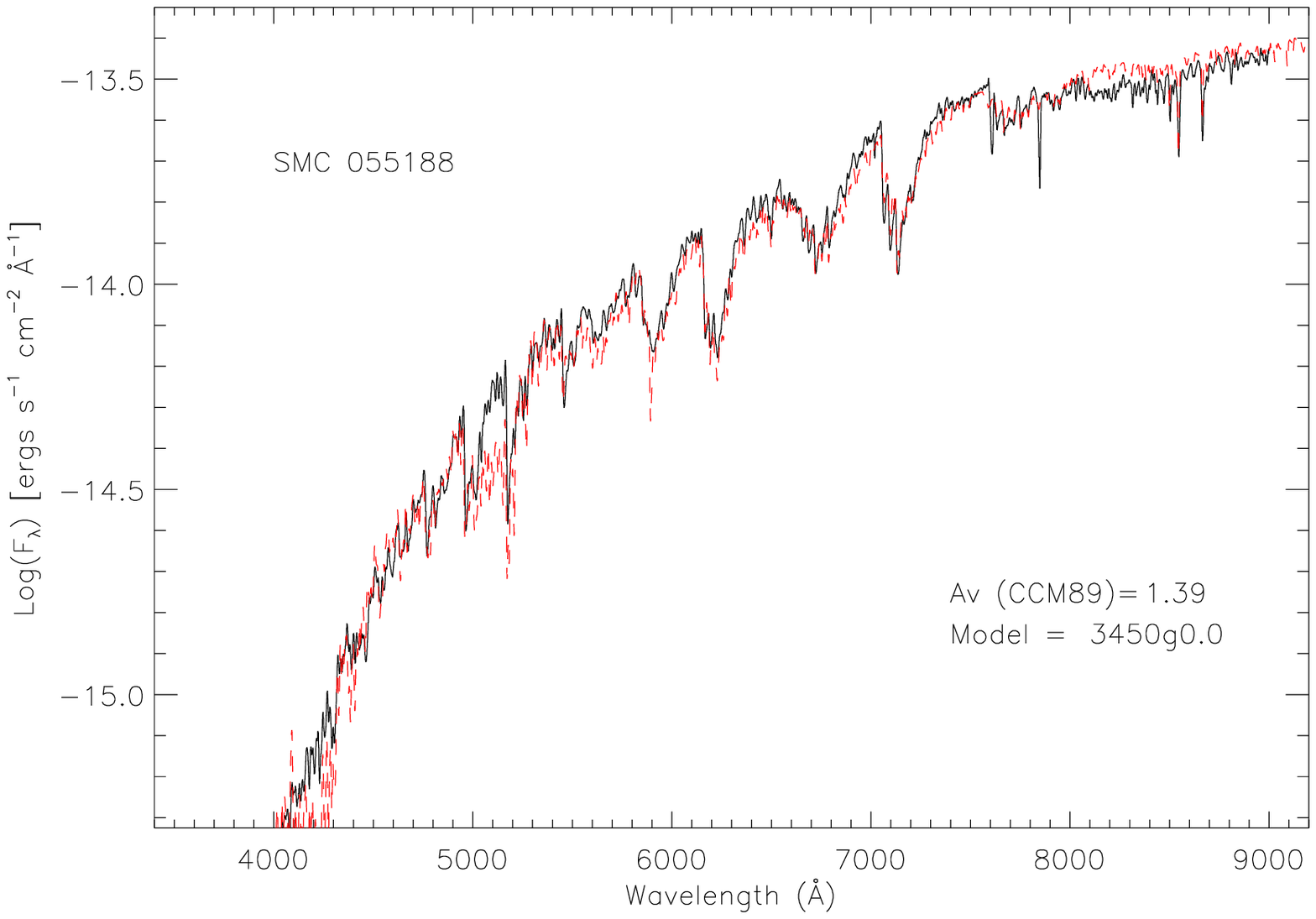}
\plotone{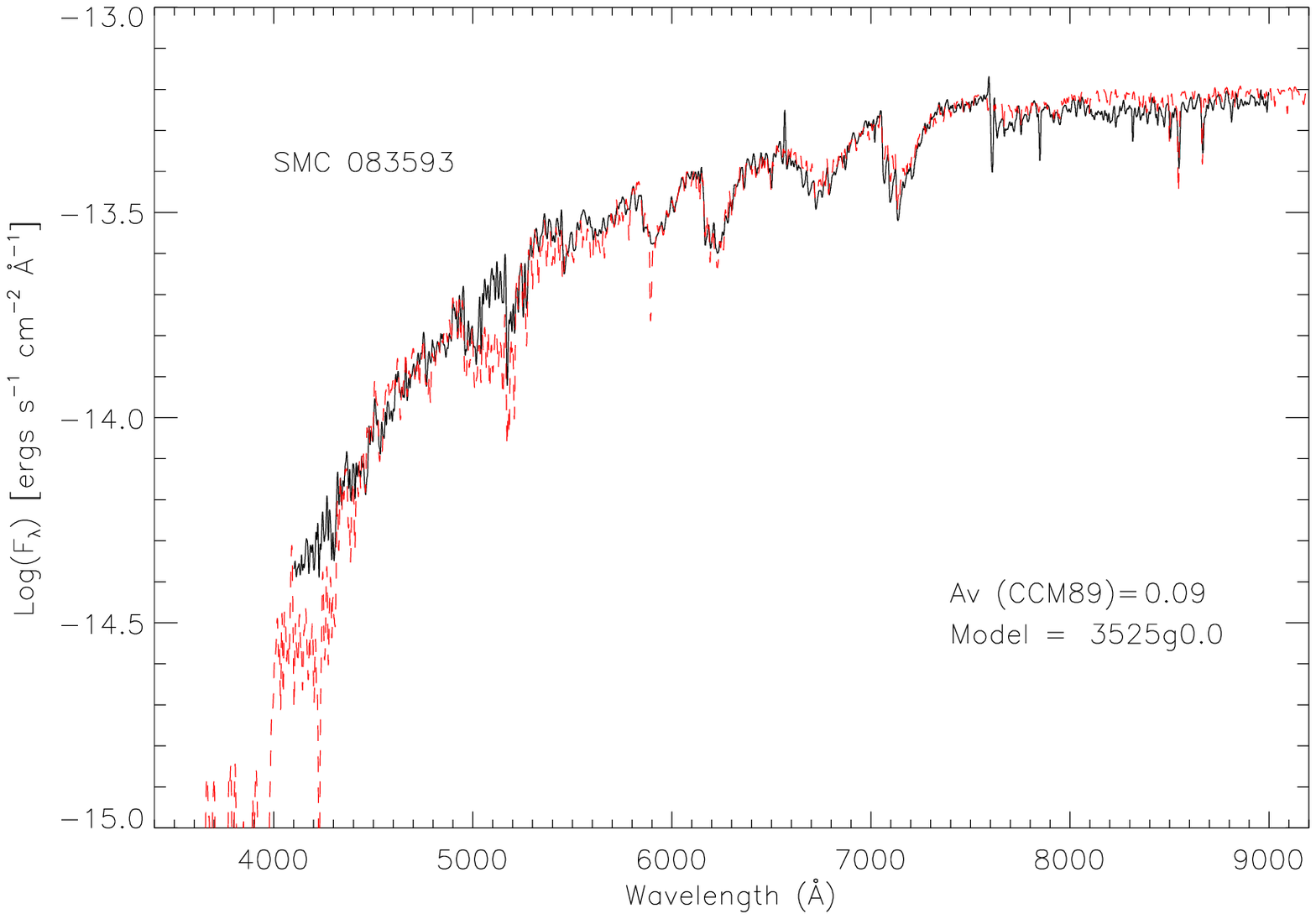}
\plotone{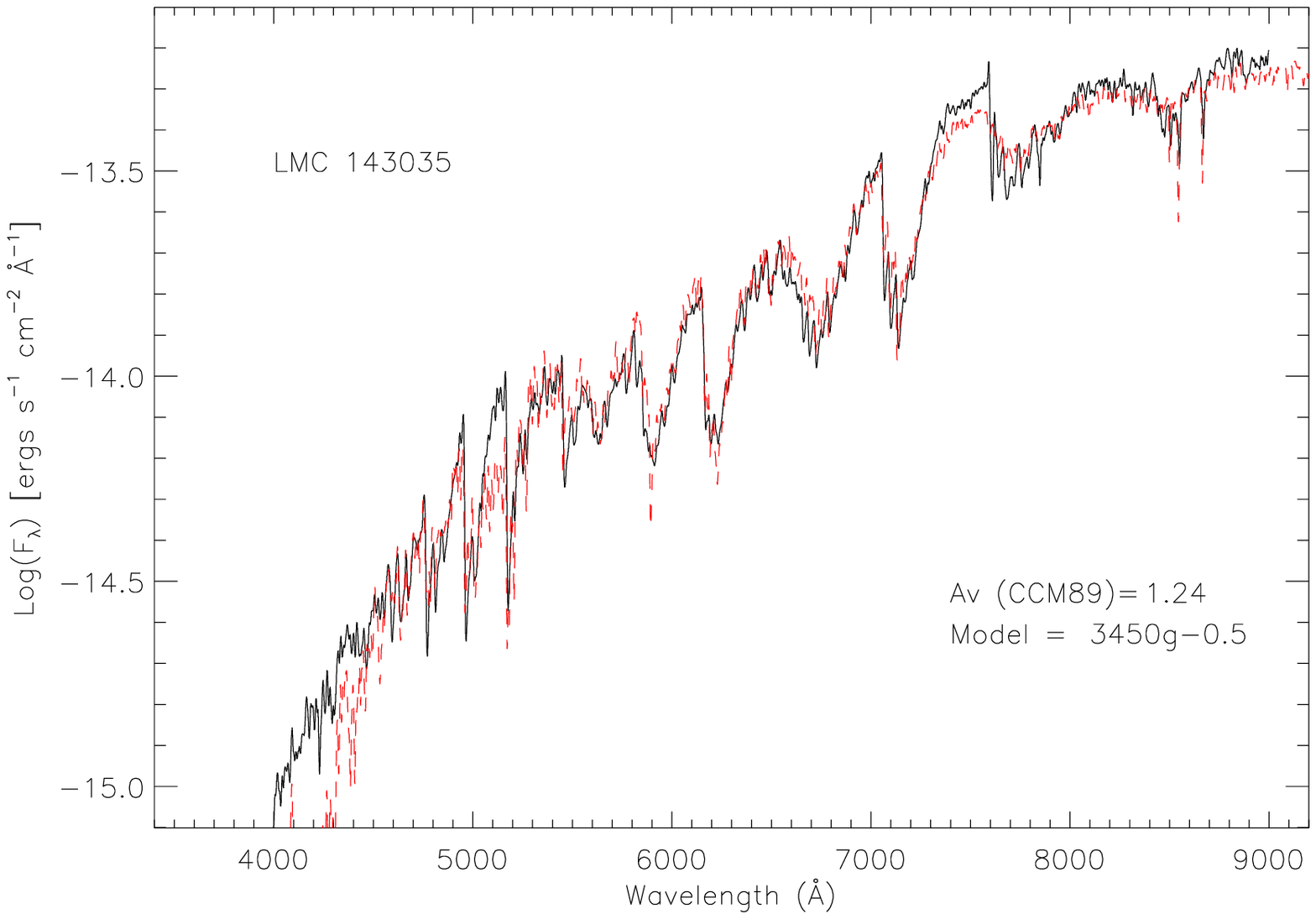}
\plotone{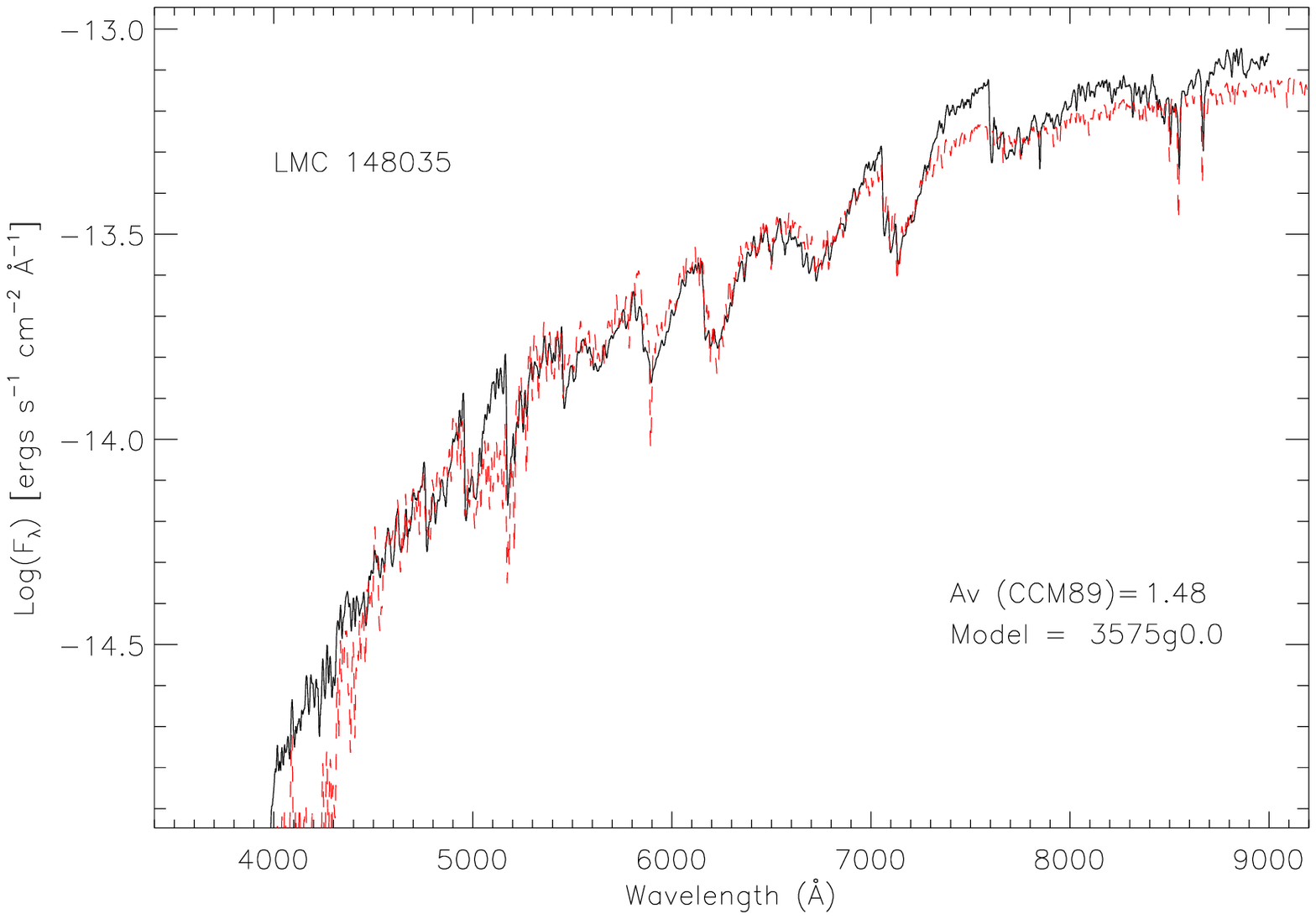}
\plotone{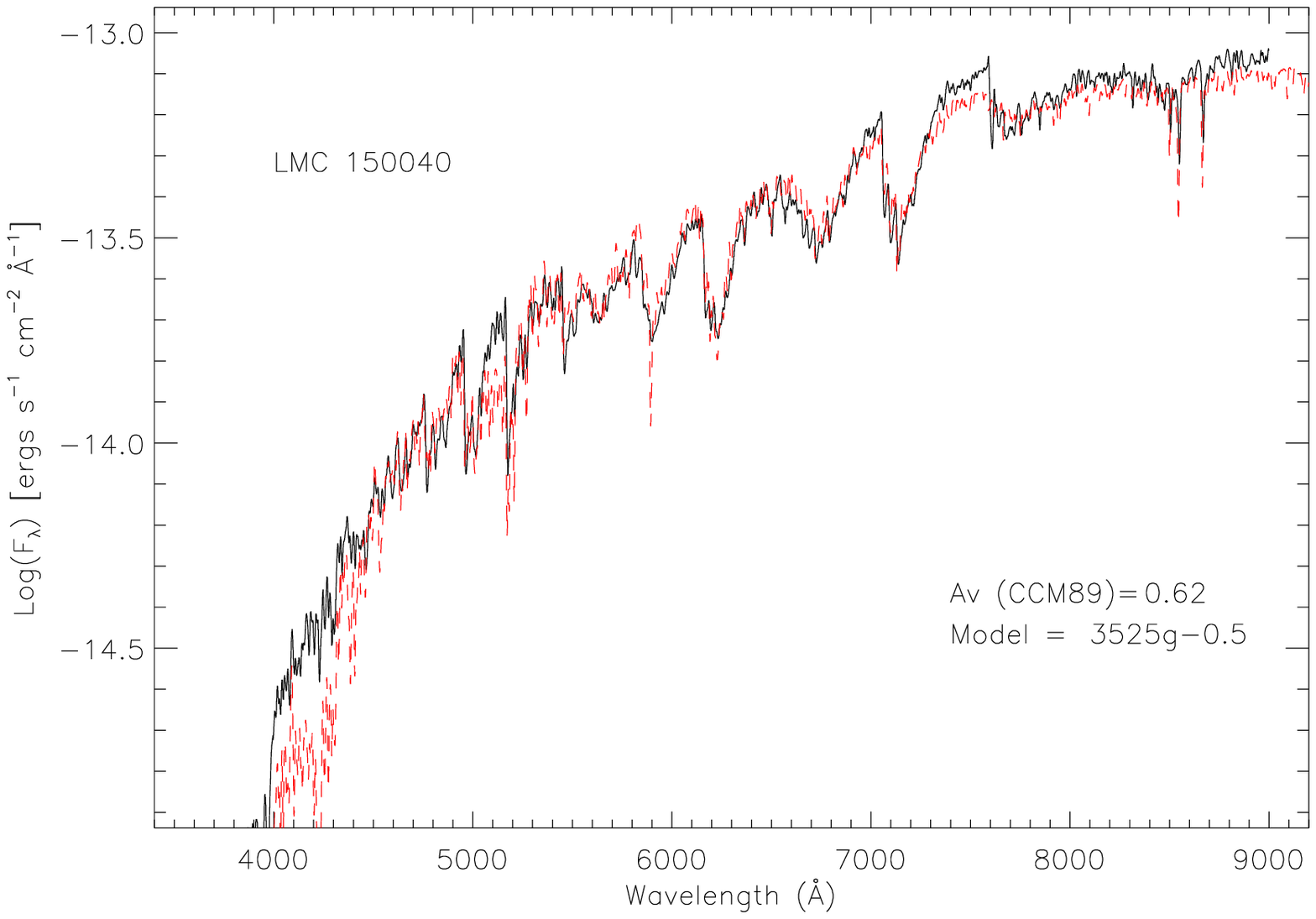}
\plotone{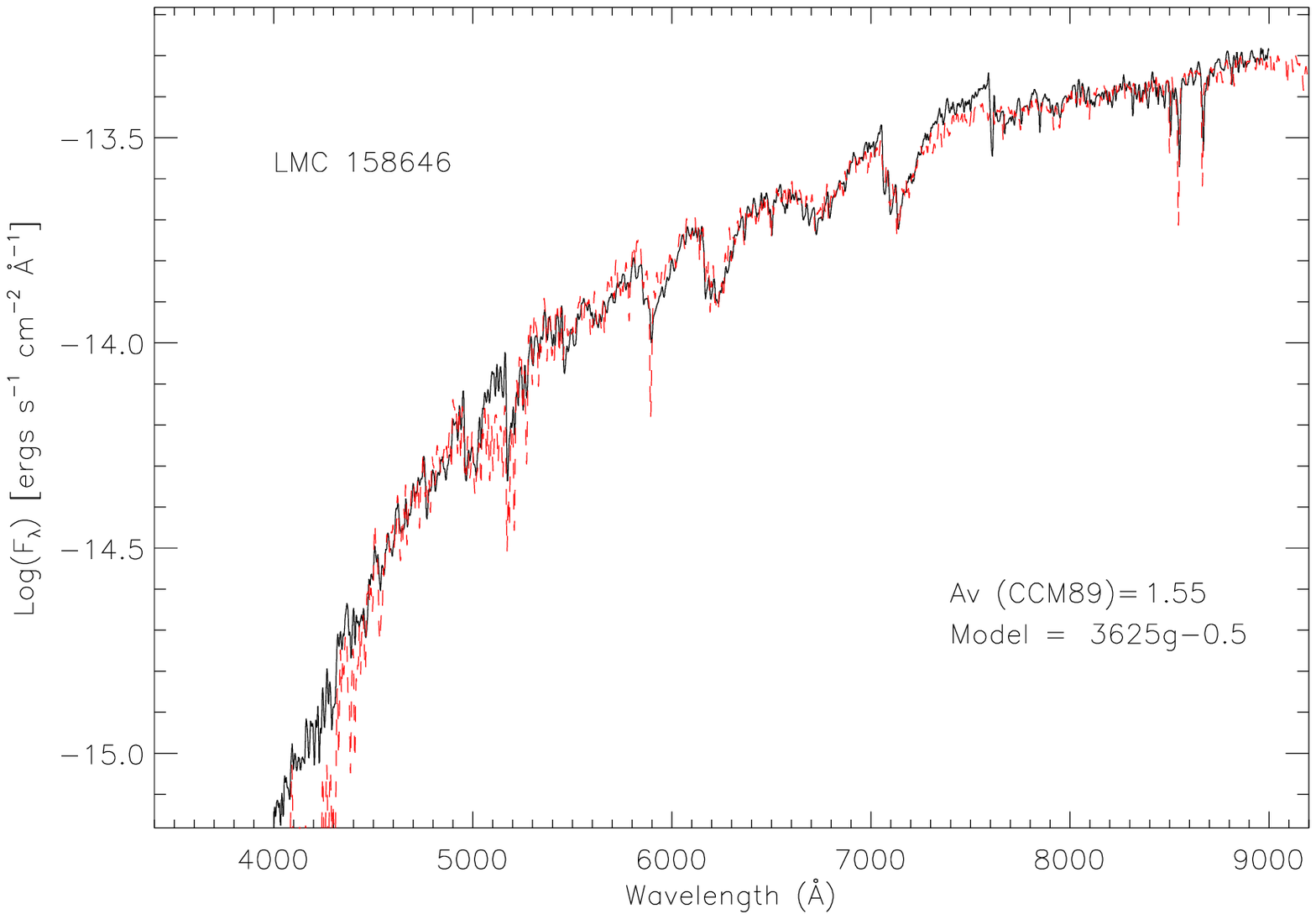}
\plotone{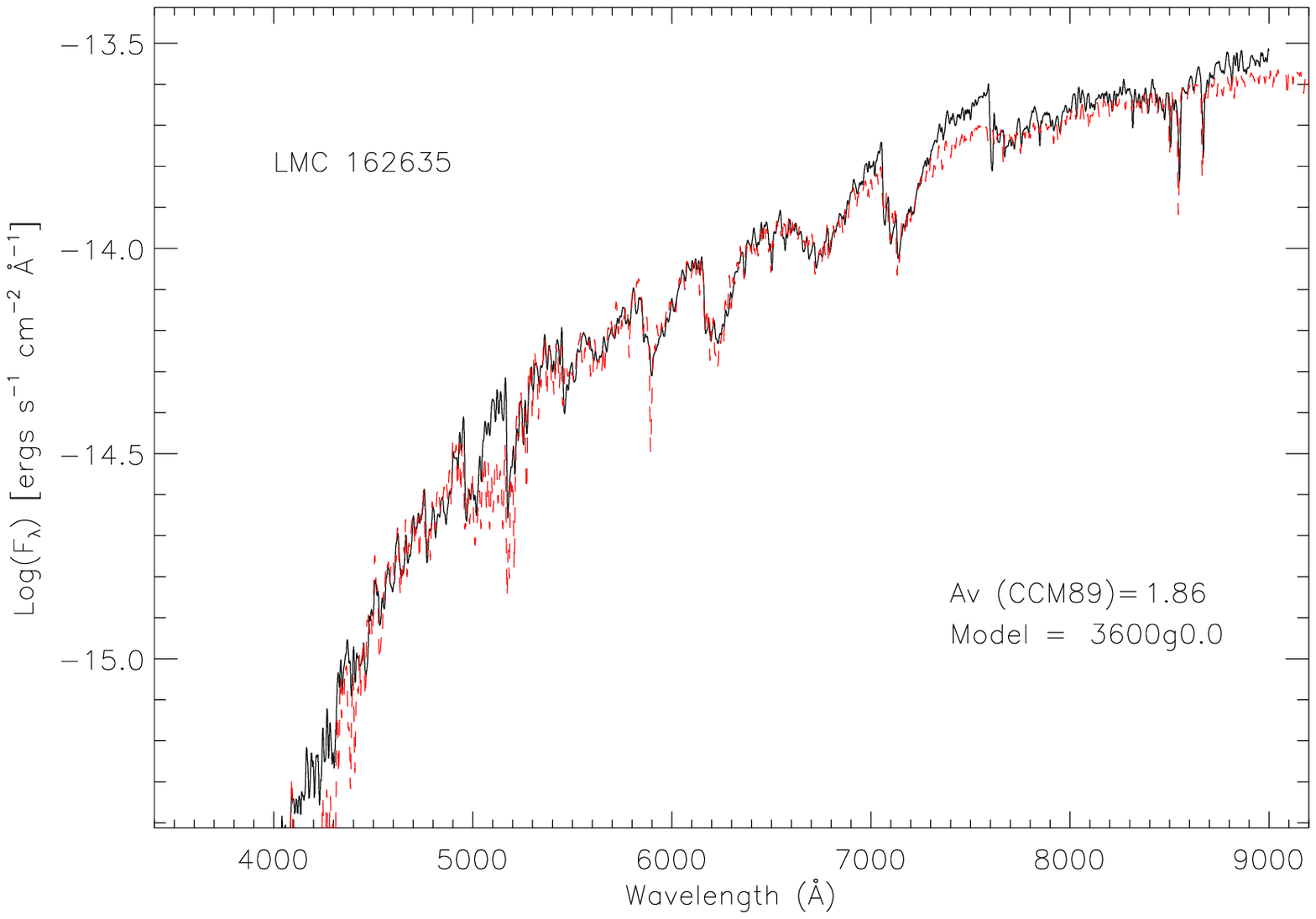}
\plotone{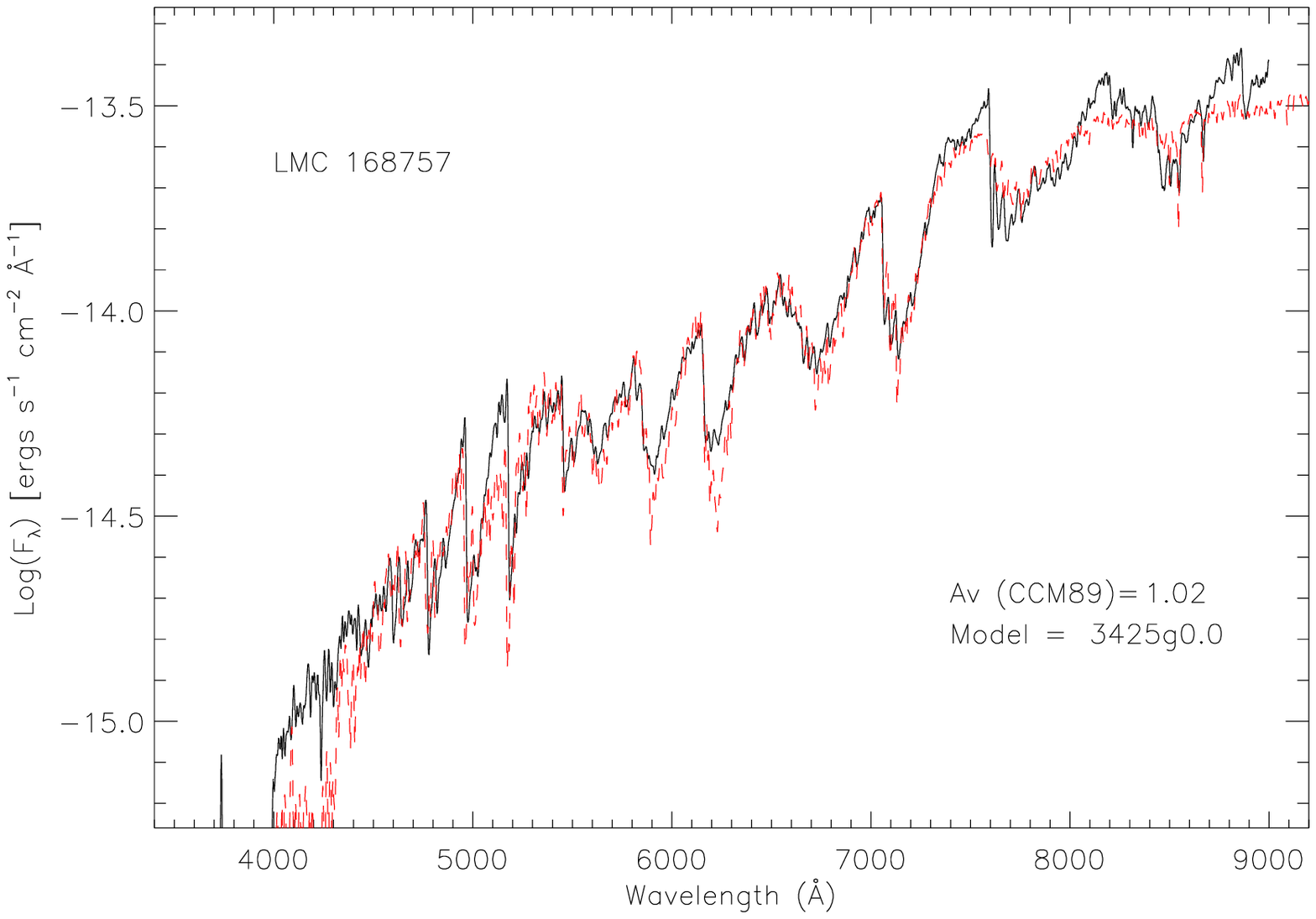}
\plotone{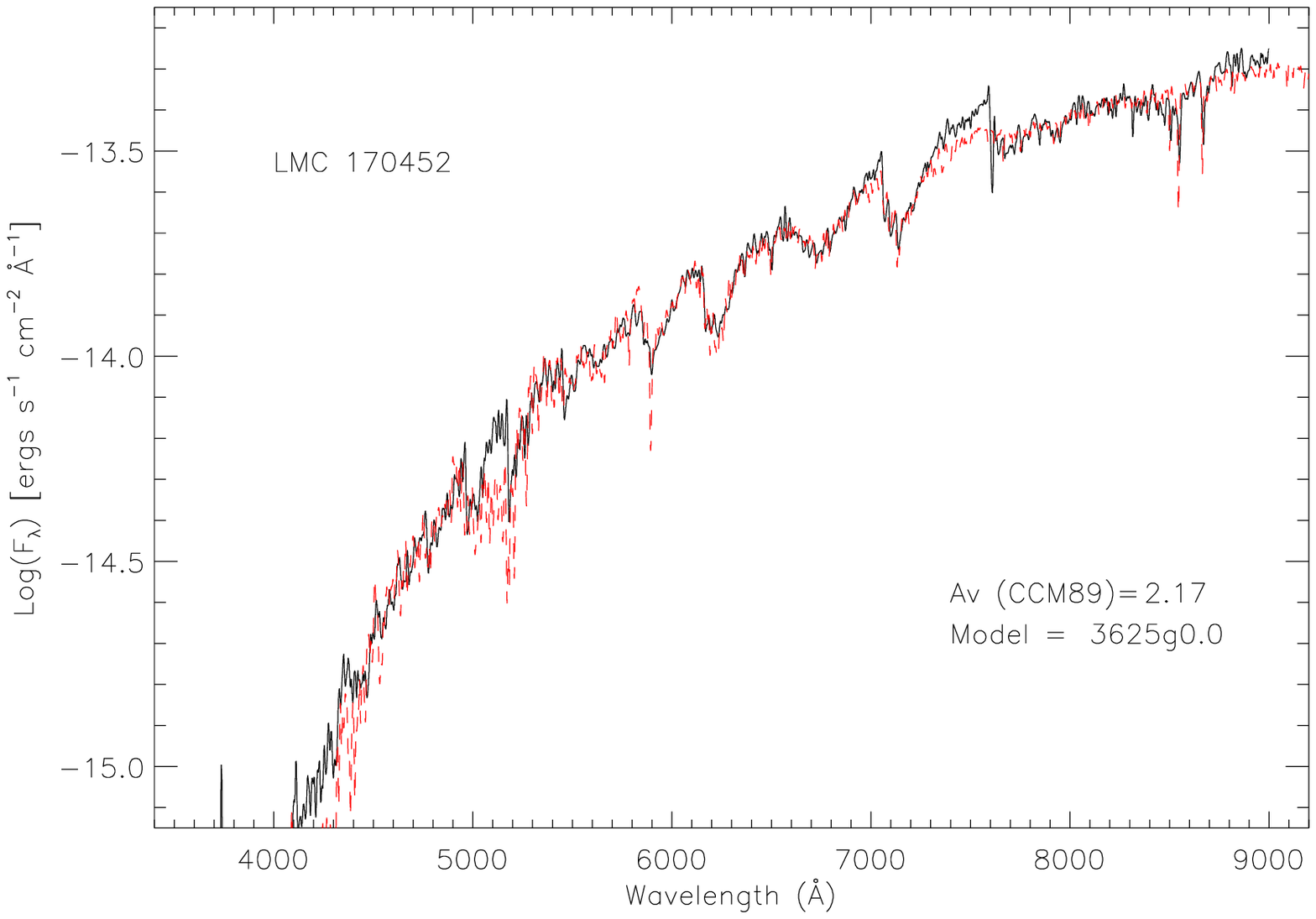}
\caption{\label{fig:fits} Fits of the MARCS stellar atmosphere models
to our spectrophotometry. The data are shown as a solid black line, while
the reddened MARCS models are shown as a dotted gray line.  In the on-line version,
the model fits are shown in red. Note the additional fits for the 2004
spectrophotometry in the cases of SMC 046662 and SMC 055188.}
\end{figure}

Our fits are shown in Figure~\ref{fig:fits}. For the purposes of scaling, we have truncated the spectra to 4000\AA.  In general the models show excellent agreement with our
observed spectral energy distributions. The excess flux in the near-UV region of the spectra is
likely due to circumstellar dust, scattering the light from the star into the line of sight; see Massey
et al.\ (2005). The synthetic spectra has to be smoothed to match the
resolution of the data. While a small mismatch can lead to discrepancies
in the comparison of atomic lines (such as the superimposing of the Mg I triplet near 5200\AA, a feature only
partially resolved in our spectra, onto TiO $\lambda$5167), the details of
the smoothing are unimportant for comparison of
the broader TiO molecular bands.

Since these stars are considerably variable in $V$ (\S~\ref{Sec-var}) we chose to
use our spectrophotometry to provide a contemporaneous measurement.
We obtained these values from our spectrophotometry following the
procedures of Bessell et al.\ (1998). We also computed contemporary $B-V$
values the same way.  
These values are listed in Table~\ref{tab:stars}.  This then allows the bolometric luminosity
we derive to be directly related to the reddening and effective temperatures we
derive, both of which may vary with time (Massey et al.\ 2007).   To derive
$M_{\rm bol}$, we first calculated $M_V$ through the simple equation
$$M_V = V - A_V - {\rm DM}$$
assuming true distance moduli (DM) of 18.9 for the SMC and 18.5 for the LMC. The bolometric correction
$BC_V$ was calculated as a function of effective temperature based on fits made to the MARCS models;
the equations are given in Section 3.2 of Paper~II.
With these quantities, $M_{\rm bol}$ is then simply
$$M_{\rm bol} = M_V + {\rm BC}_V$$
 From the bolometric luminosity we can derive the ratio $L/L_{\odot}$ by
$$(L/L_{\odot}) = 10^{(M_{\rm bol} - 4.74)/(-2.5)}$$
and use the luminosity-radius-temperature relation:
$$R/R_{\odot} = (L/L_{\odot})^{0.5} (T_{\rm eff}/5770)^{-2}$$
to obtain the stellar radii.

\subsection{Alternative Method Using K-band Photometry}
\label{Sec-bb}

In Papers I and II we found it was very useful to have some (nearly) independent
check on our results by using the existing $K$-band photometry.   In general,
the variability at $K$ is less than that at $V$ (Josselin et al.\ 2000); we find
below (\S~\ref{Sec-var}) that this is true for these stars as well.

In order to derive (nearly)
independent values of $T_{\rm eff}$ and $M_{\rm bol}$ from the K-band
photometry,  we dereddened the $V-K$ values from Table~\ref{tab:stars} using
$(V-K)_0 = V - K - 0.88A_V$, where the numerical value was derived in Paper II
(and is in agreement with that of Schlegel et al.\ 1998), and the values for
$A_V$ are taken from Table~\ref{tab:results}.

We derive the effective temperatures from $(V-K)_0$ using relations
derived from the models and given in Section 3.3.1 of Paper~II.

To derive the bolometric luminosity using these new temperatures,
we first calculated $M_K$ by:
$$M_K = K - A_K -{\rm DM}$$
where $A_K = A_V \times 0.12$.  We then derived the
bolometric corrections to the K-band (BC$_K$) using
the K-band effective temperatures and the relations
given in Paper~II.
The bolometric luminosity is then simply:
$$M_{\rm bol} = M_K + {\rm BC}_K$$
and the stellar radius follows as shown in \S~\ref{Sec-spec}. The values
for BC$_K$ came from adopting the $(V-K)_0$ temperatures and
using a relation derived from the MARCS models in Paper II.

We include these values in Table~\ref{tab:results}, and they show good general
agreement with the values obtained from spectrophotometry. Systematically, the
K-band temperatures yield a median difference of $-$200 K in temperature
for the 4 SMC stars and $-$96 K in temperature
for the 7 LMC stars, in the sense that the K-band values are larger.
These systematic differences are similar to those found from
the larger sample considered in Paper~II, where we
found median differences of $-$170~K and $-$105~K between the K-band photometry
and spectral fitting for the SMC and LMC, respectively.  The median K-band
luminosities are lower, 0.41~mag for the SMC, and 0.18~mag for the LMC.
In Paper~II we attributed the differences to the inherent limitations of the 1-D 
atmosphere models.

\subsection{Variability}
\label{Sec-var}

\subsubsection{Photometric Variability}
Many of the objects in our sample demonstrate large variability in their $V$ magnitudes. Photometry was obtained from the All Sky Automated Survey (ASAS) project (Pojmanski 2002).
To this, we added some additional data of our own.  First, we used the CCD photometry
of Massey (2002) and obtained individual measurements for each of our stars,
rather than the averages given in that paper.  These new values are given
in Table~\ref{tab:phot}.  Included as well are the values we derive from our
spectrophotometry, including that of the 2004 observing run described in
Paper II. We also list the extinction $A_V$ we derive from our spectral
fitting, in order to supplement the limited color information available. The photometry derived from the spectrophotometry
was derived using the $V$ band curve of Bessell (1990) and the zero-points
given by Bessell et al.\ (1998).

\begin{figure}
\epsscale{0.3}
\plotone{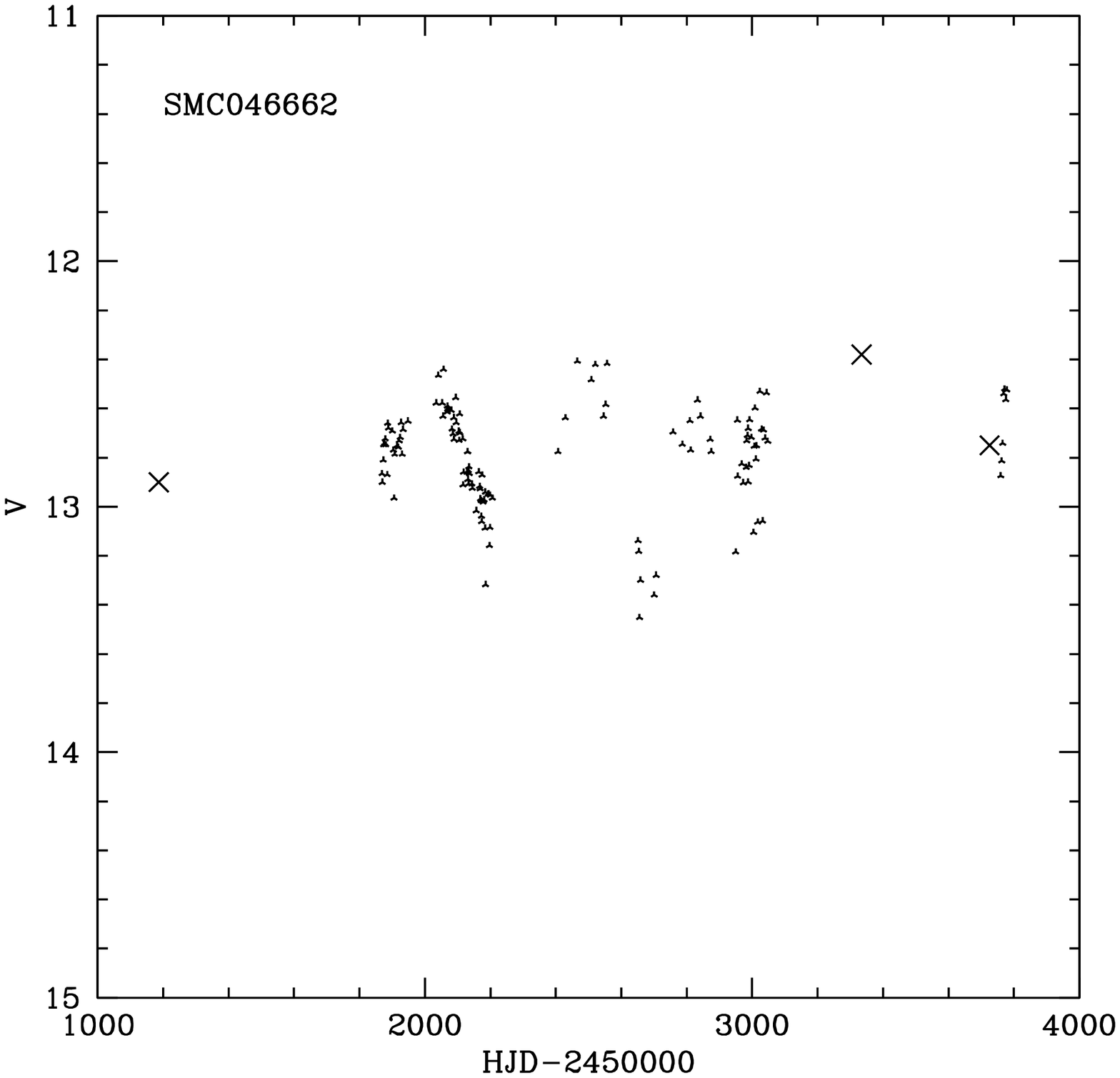}
\plotone{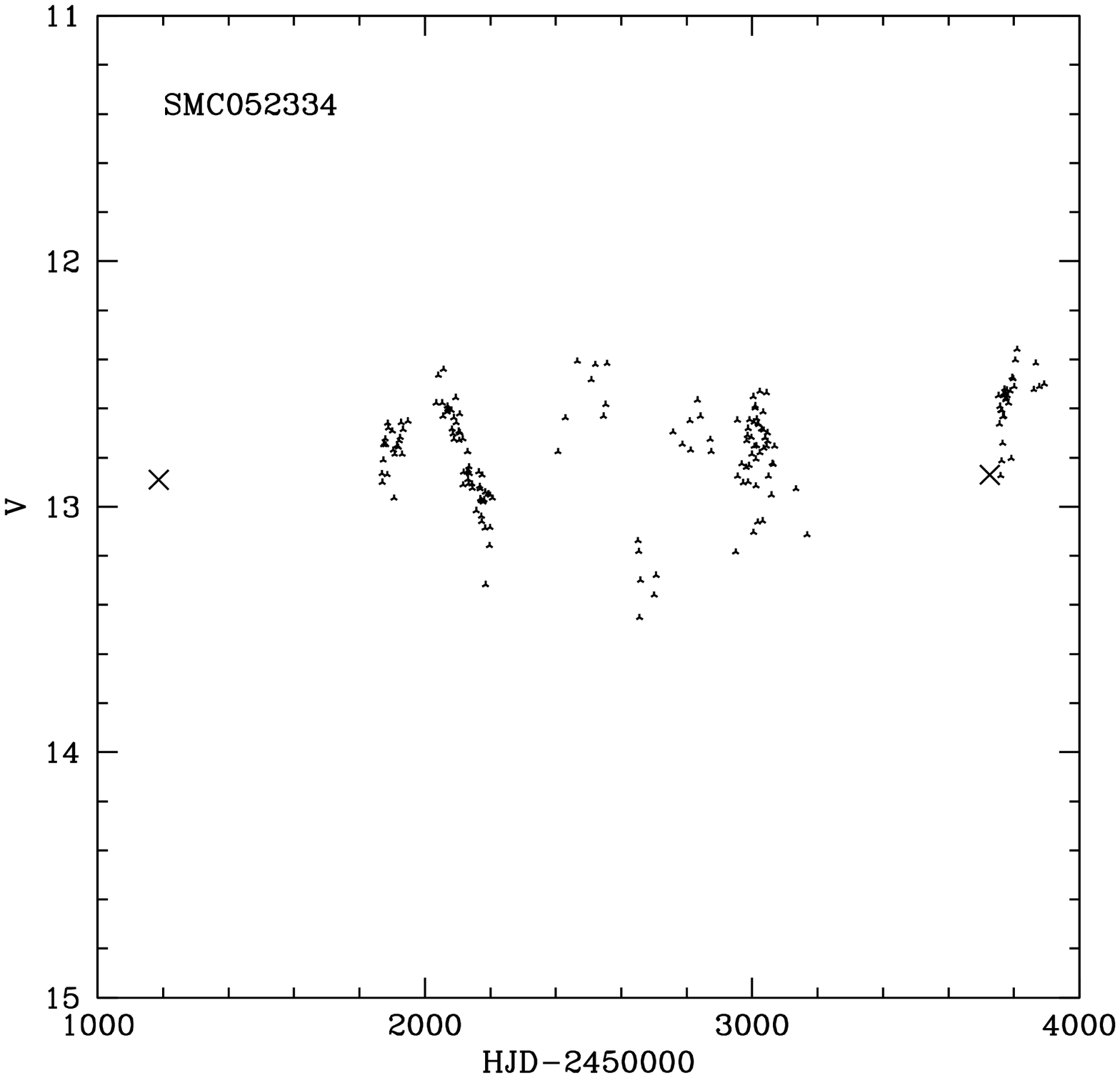}
\plotone{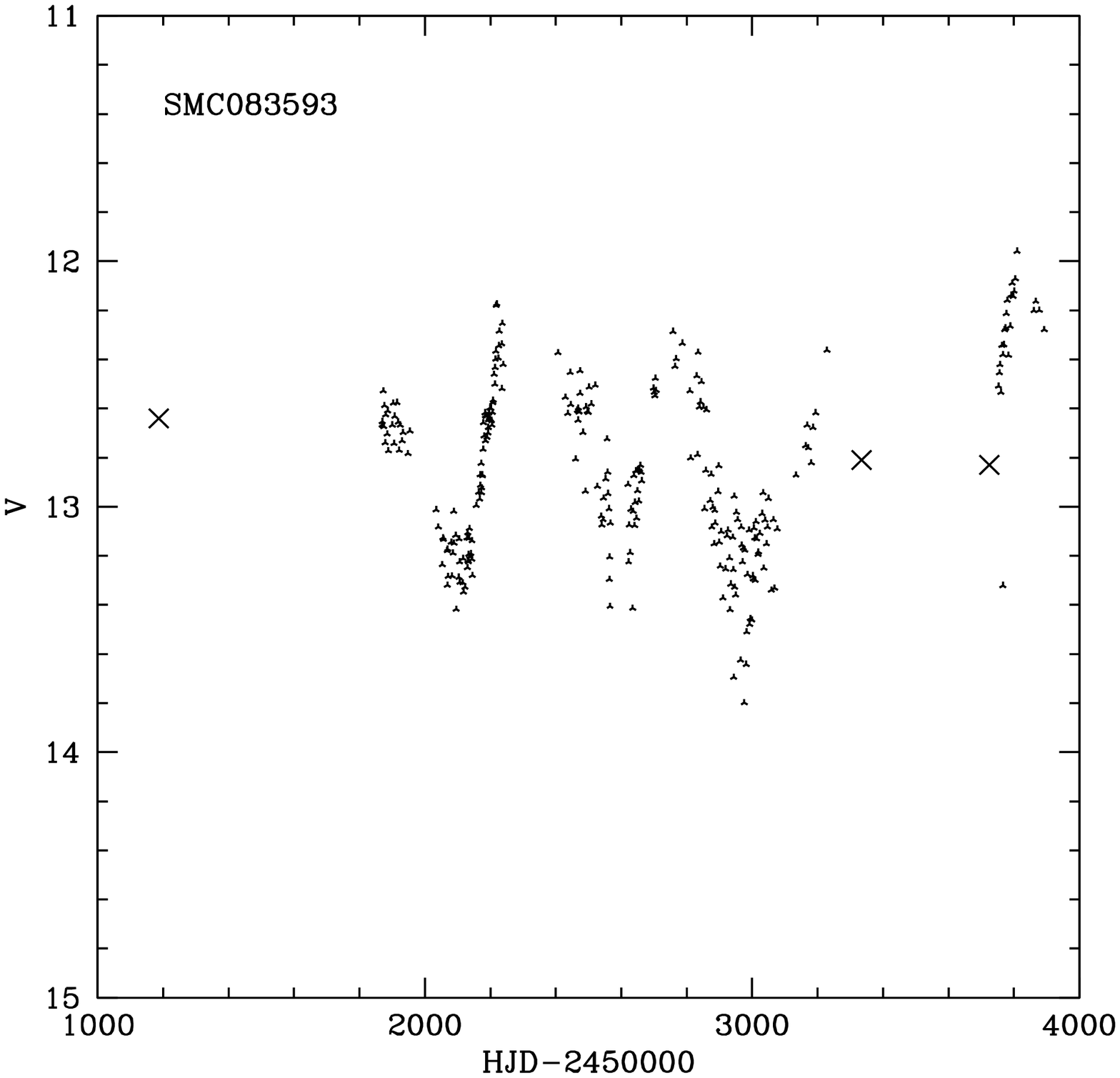}
\plotone{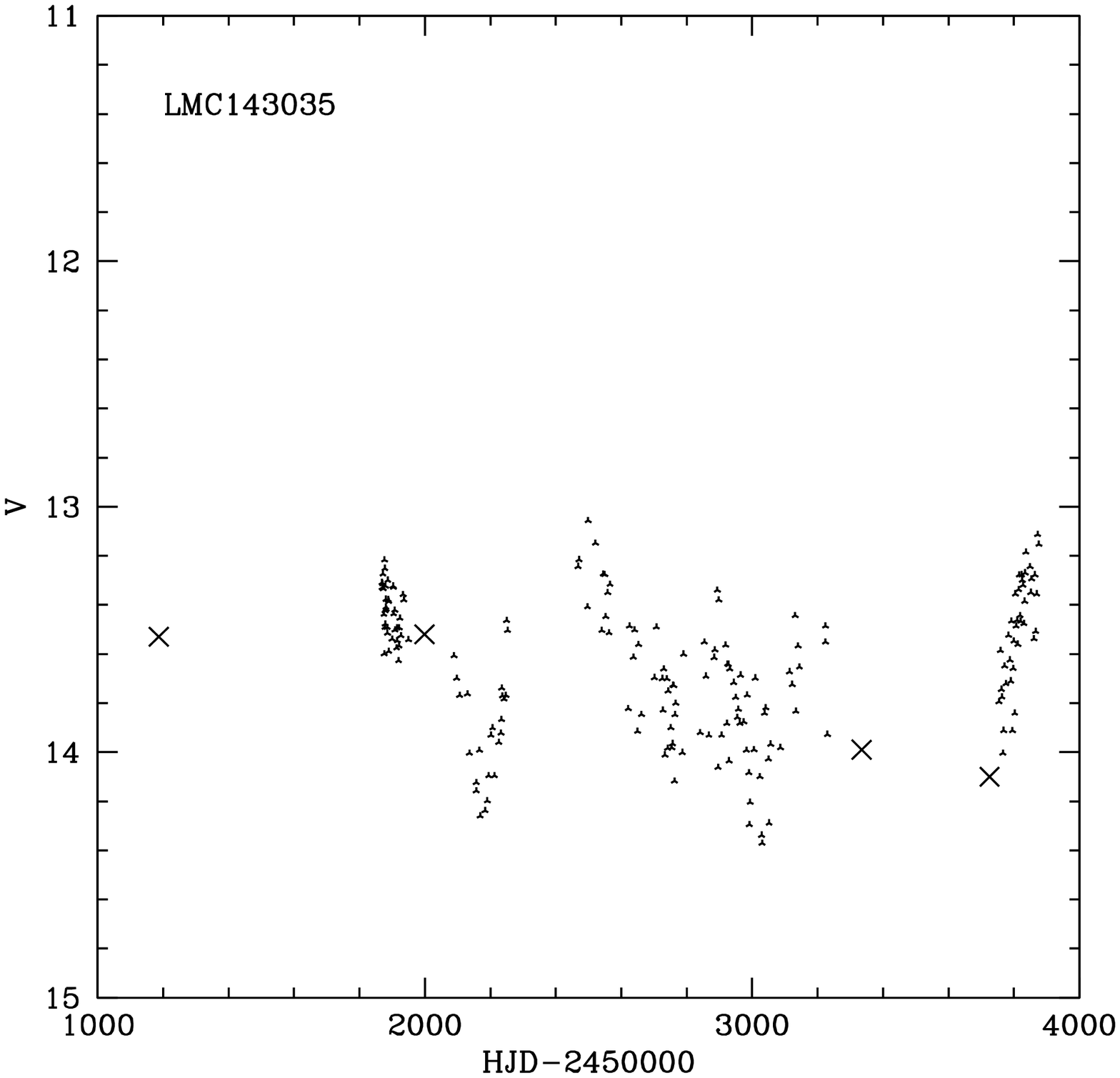}
\plotone{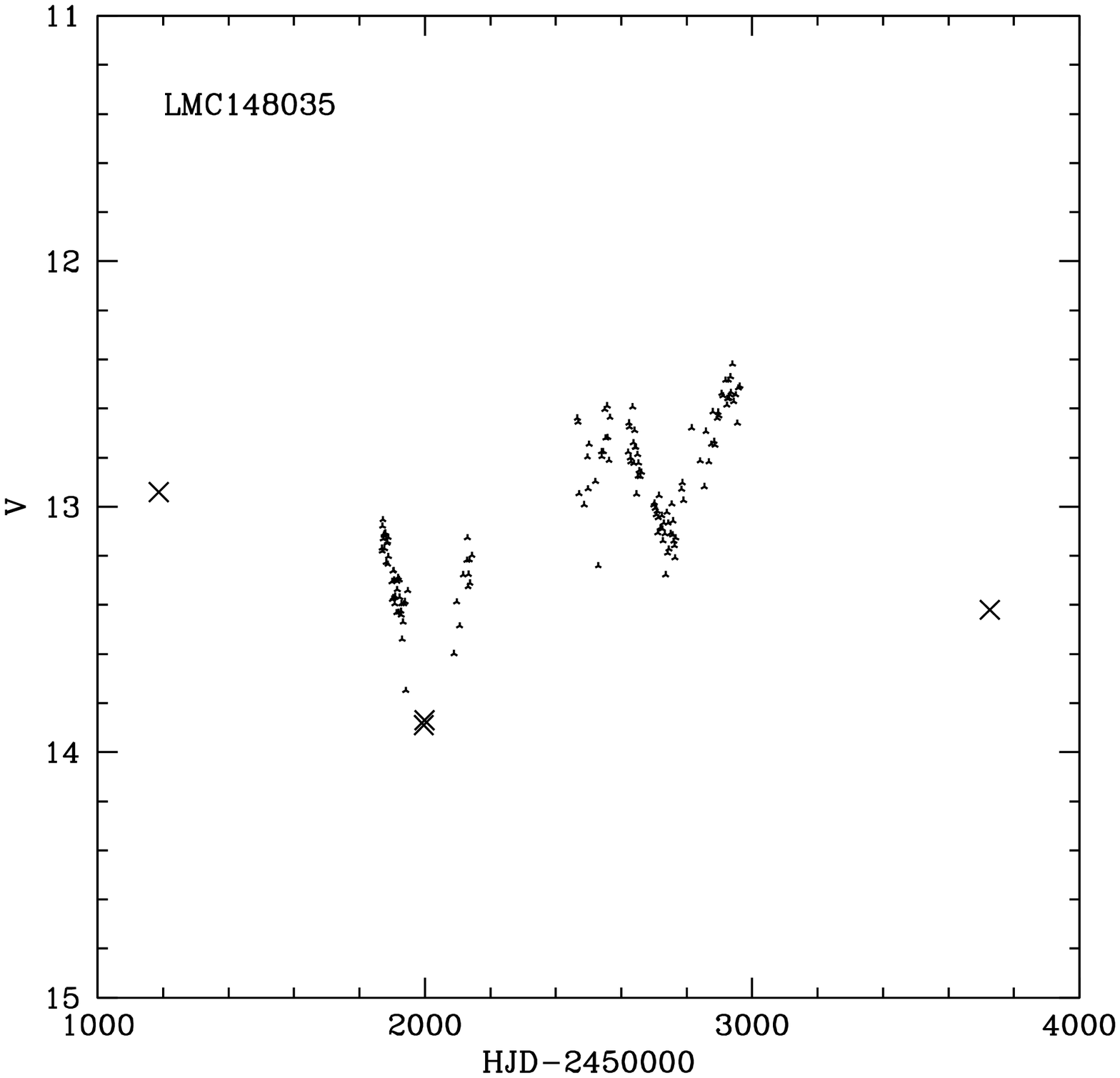}
\plotone{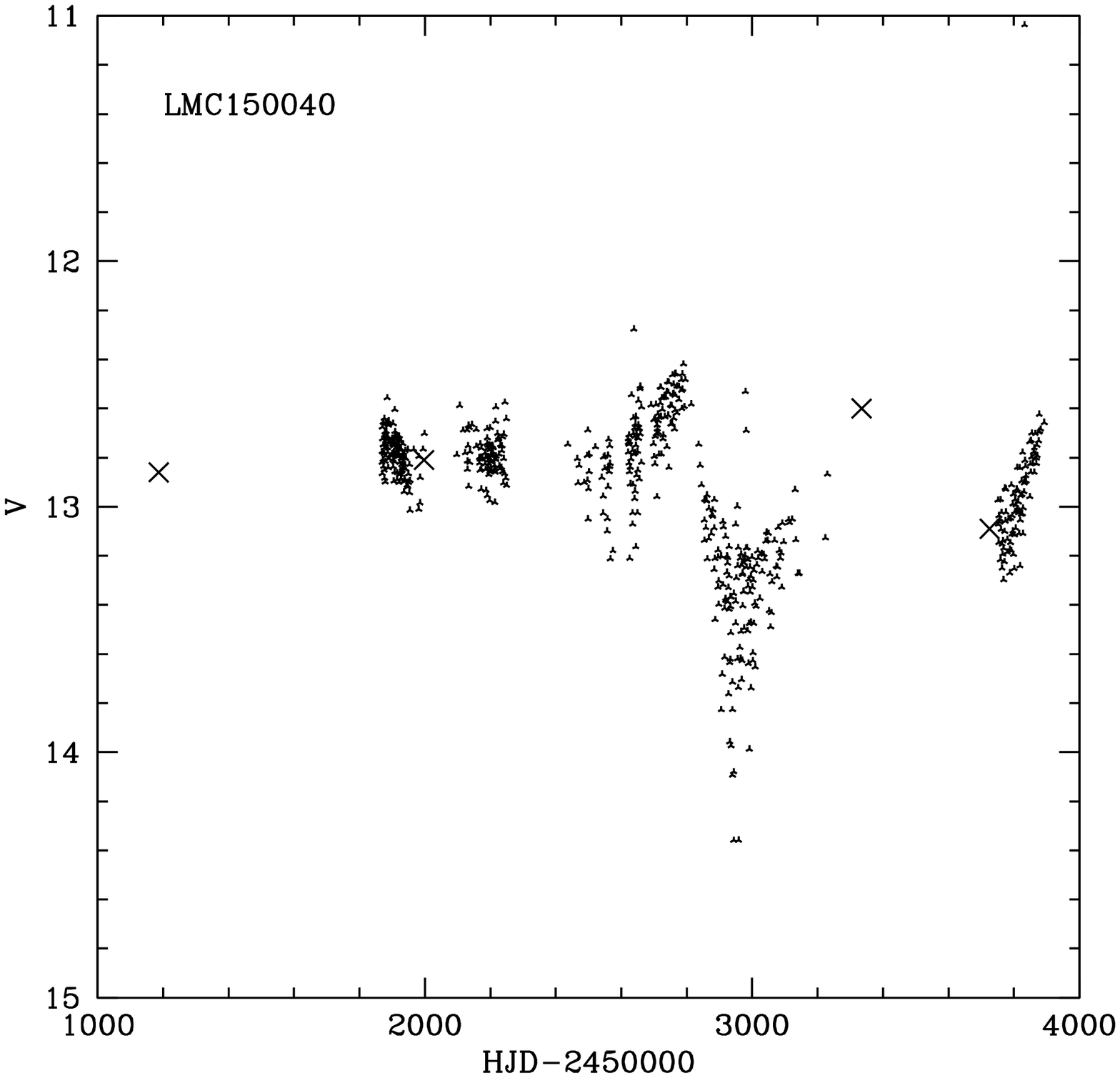}
\plotone{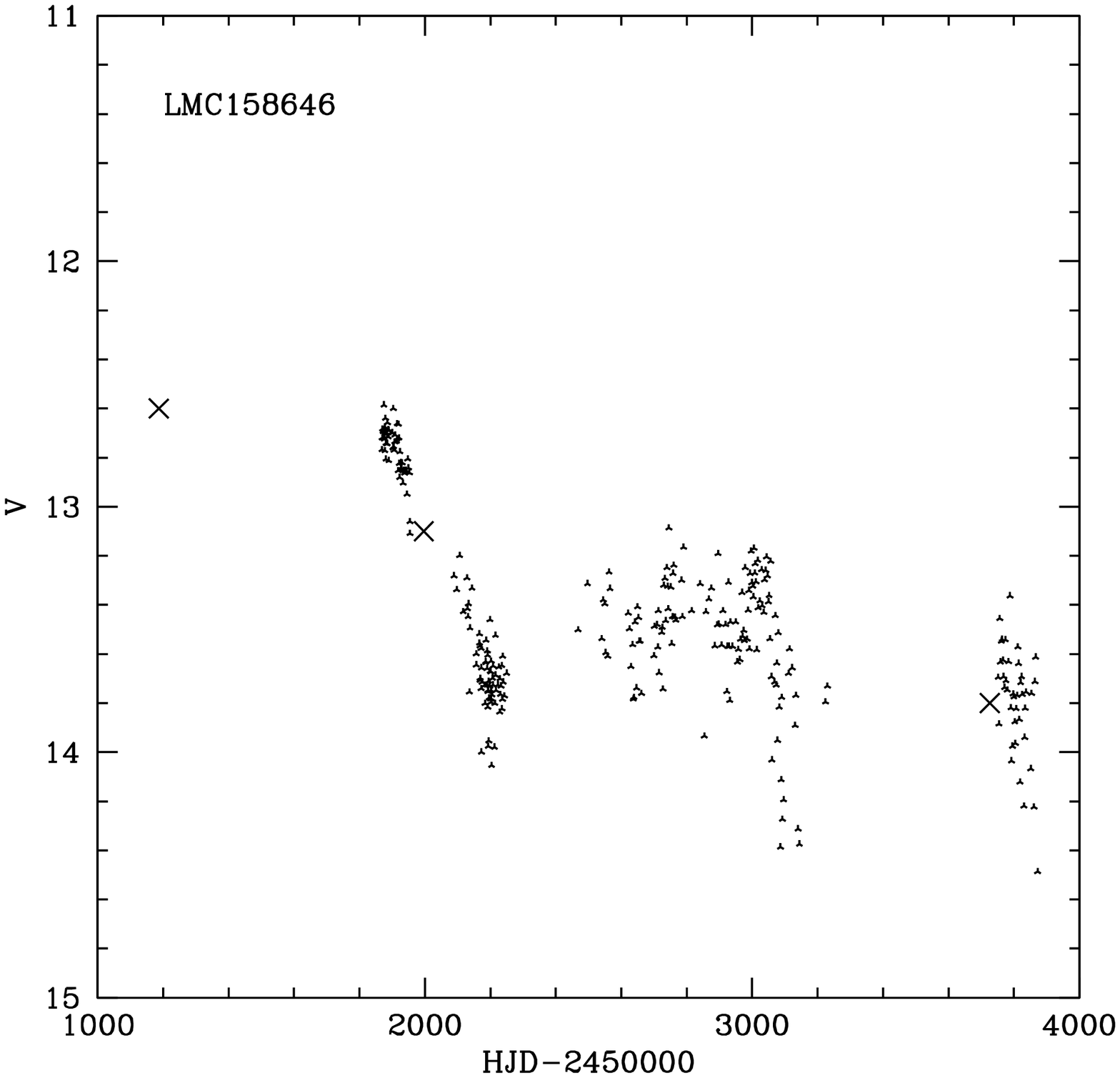}
\plotone{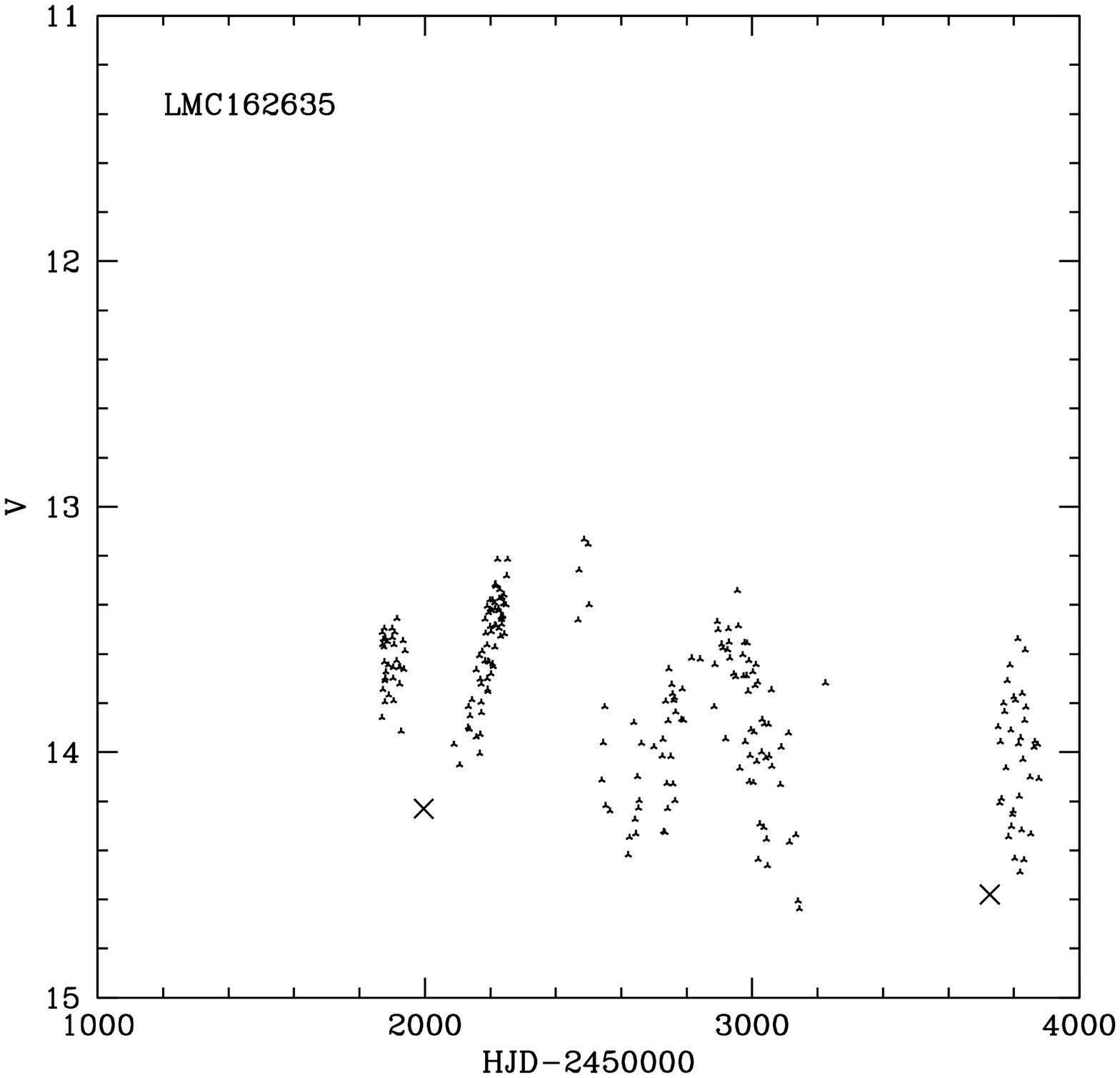}
\plotone{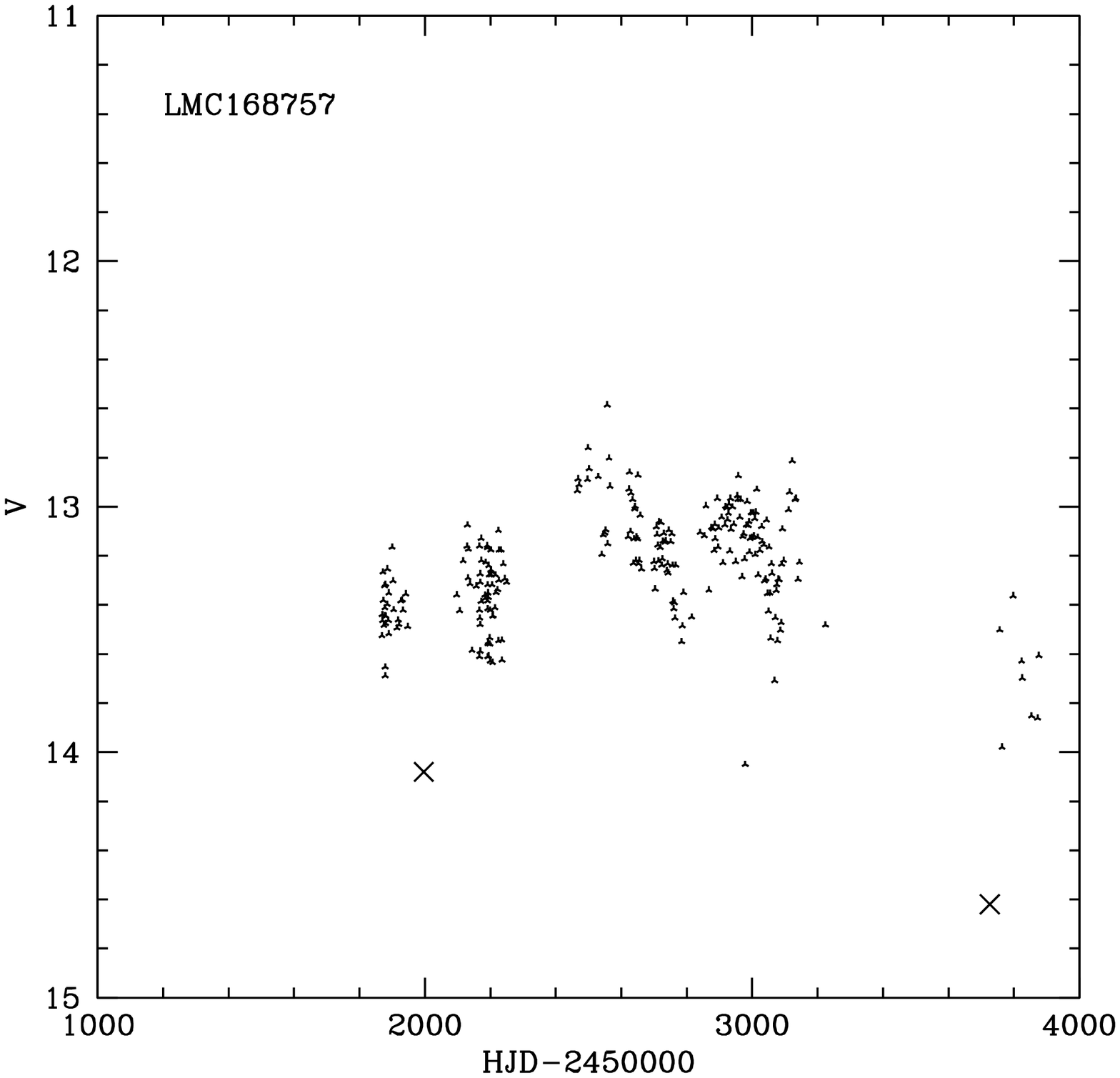}
\caption{\label{fig:phot}  Photometric variability. We show the photometry of nine of our stars, where
the small points come from the ASAS photometry, and the large points come
from Table~\ref{tab:phot}. Very few data points were  available for SMC 055188 and LMC 170452, and these stars are not included in the figure, although we argue in the
text that these two stars also show large photometric variability.}
\end{figure}

The light curves are given in 
Fig.~\ref{fig:phot} for nine of our eleven objects
(SMC 055188 and LMC 170452 lacked sufficient ASAS data). While normal RSGs
are known to be variable in $V$ (Josselin et al.\ 2001), each of the late-type
stars described here show larger variations. The maximum changes we see
for the stars discussed here average to a $\Delta V$ of 1.3 for the SMC sample
and 1.6 for the LMC sample. For comparison, the ASAS photometry of seventy
SMC and LMC RSGs studied in Paper II average to a $\Delta V$ of 0.9 mag.
Many of the late-type RSG lightcurves also seem to suggest some level of
quasi-periodicity.

While SMC 055188 and LMC 170452 are not included in Fig.~\ref{fig:phot}, our
observations in Table~\ref{tab:phot} show that they are also quite variable in
$V$. In addition, NOMAD (Zacharias et al.\ 2004) gives $V$ magnitudes of 13.82
and 14.91 for these stars, respectively. So, it appears that these stars show
$V$-band variability of 1.2-1.3 mag.

In contrast, we expected that the variability at $K$ would be less; at least, this is typical
of other RSGs (Josselin et al.\ 2000).  We checked the 2MASS values
against DENIS (Kimeswenger et al.\ 2004),
and found much lower
variability, with an average difference at $K$ between the two surveys
of 0.08 mag.

Finally, we note that for four of the stars,
the $B-V$ photometry we derive from our (2005) spectrophotometry  
differs significantly
($>0.1$ mag) from the 1999/2000 values of Massey (2002).  We list both  
the new and old values
in Table~\ref{tab:phot}.  In general the agreement is good, but for  
four of our stars (SMC 083593, LMC 148035,
LMC 162635, and LMC 170452) we see differences of several tenths in  
$B-V$.  As discussed in our
study of HV 11423 (Massey et al.\ 2007) $B-V$ is not very sensitive  
to effective temperatures for RSGs.
Instead we suggest that the variations we see in $B-V$ are indicative  
of changes in the amount of circumstellar
dust causing differences in the reddening.  (As discussed by Massey  
et al.\ 2005, dust around
RSGs {\it should}, and apparently does, result in a significant  
amount of circumstellar extinction; in the extreme
cases in the Milky Way, amounting to several magnitudes.)  We saw  
similar changes in $B-V$ for HV 11423,
and suggests episodic dust ejection on the time scale of a few years,  
consistent with the study of Danchi et
al. (1994).

Note that these changes in $B-V$ are not correlated in some simple way with
changes in $V$: a redder color does not necessarily mean that the star is
fainter. This demonstrates that the $V$-band variability is {\it not} simply
caused by changes in the amount of circumstellar extinction, but that
rather real changes in the star (such as effective temperature) are
responsible for the change in $V$. For HV 11423 we argued that these
physical changes were also triggering bursts of enhanced dust production,
further complicating the light curve.

\subsubsection{Spectral Variability}

We were intrigued by the large discrepancies in spectral subtypes assigned to some of our stars
by previous studies and ourselves (Table~\ref{tab:stars}).  Although assigning spectral types is a
little subjective, we did not see many such differences in the comparison of Massey \& Olsen (2003)
types with ours in Paper~II.  Yet, spectral variability of a type or more is virtually unknown for RSGs.
The notable past exception is the SMC RSG star HV~11423, recently found by Massey et al.\ (2007) to have varied several times between K0-1~I and M4.5-5~I in the past several years.  At its coolest,
it is the latest type supergiant in the SMC.  HV~11423 shows large photometric variability at $V$, due in part
to the substantial change in effective temperature (while holding relatively constant in bolometric luminosity), and
in part due to the increased circumstellar extinction, presumably from outbursts of dust formation resulting from mass loss.

We were therefore interested to follow up the question about whether or not the stars discussed
here are truly variable or not in spectral type. We have included spectral
types for these stars in Table~\ref{tab:phot} from Massey \& Olsen (2003) and this study,
as well as types from our 2004 data.  While in most cases we do not have
sufficient wavelength coverage to determine physical properties, the data
we do have is sufficient to determine accurate spectral types, and therefore
useful for evaluating the spectral variability of these stars over the past
few years.

\begin{figure}
\epsscale{0.4}
\plotone{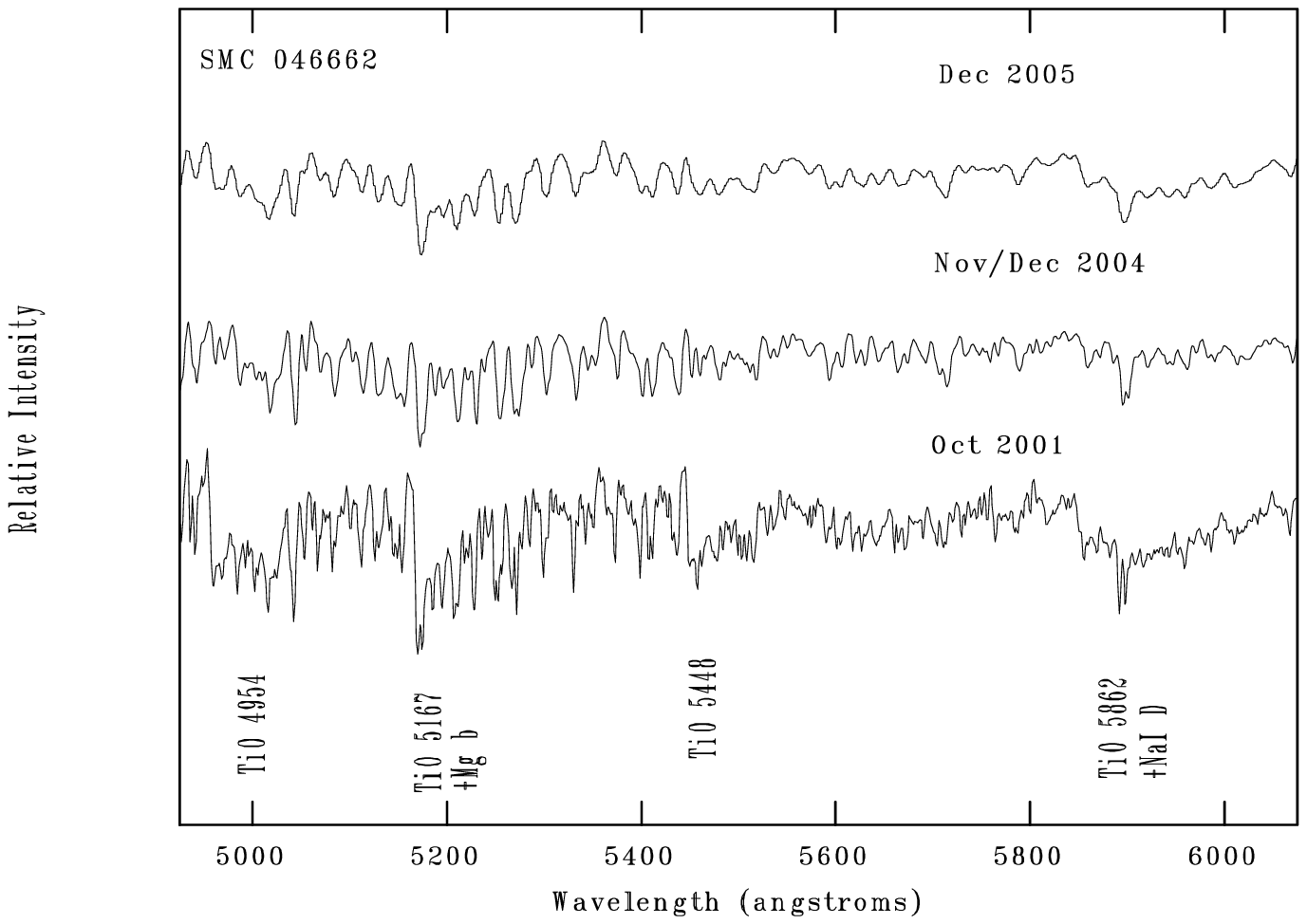}
\plotone{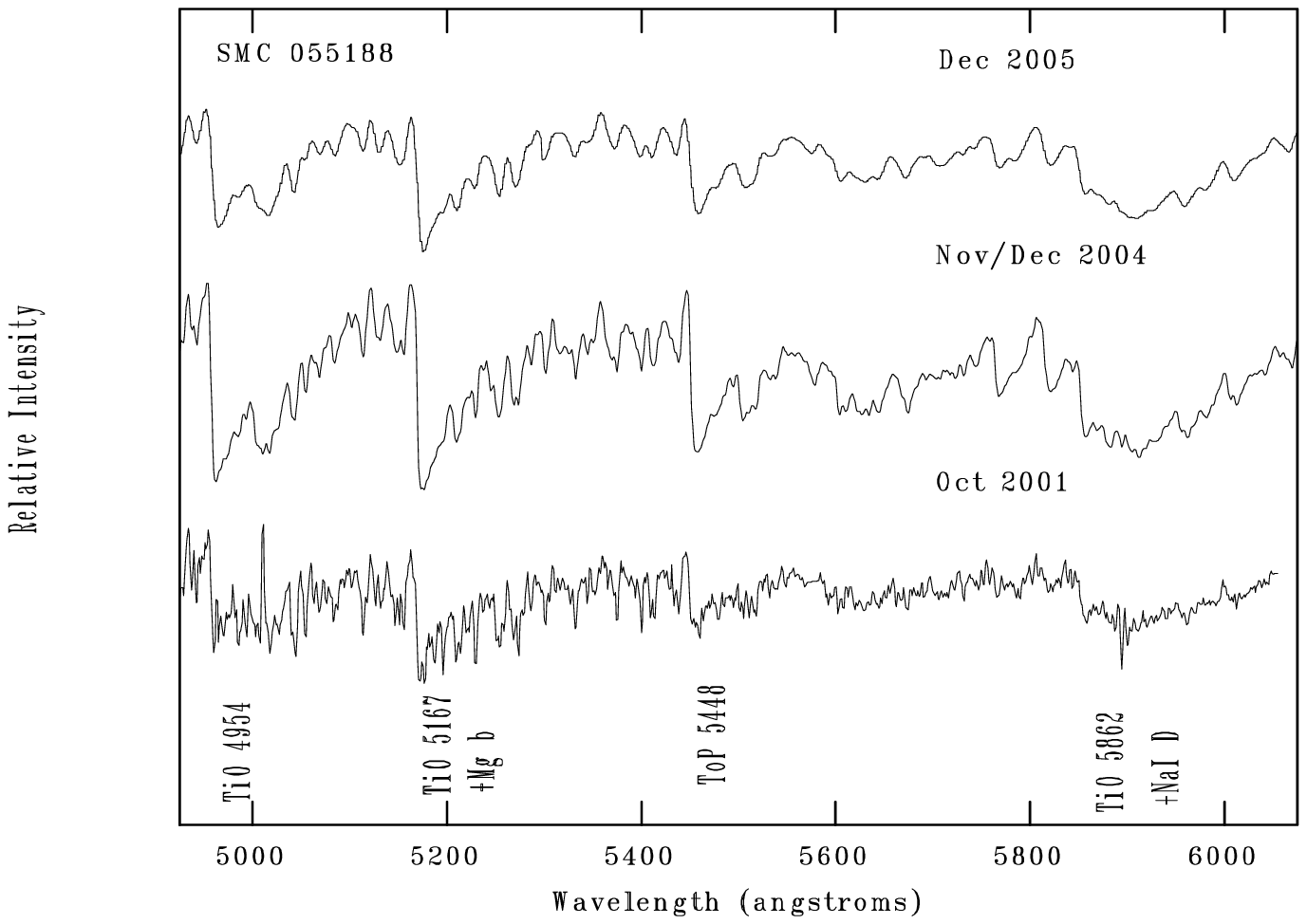}
\plotone{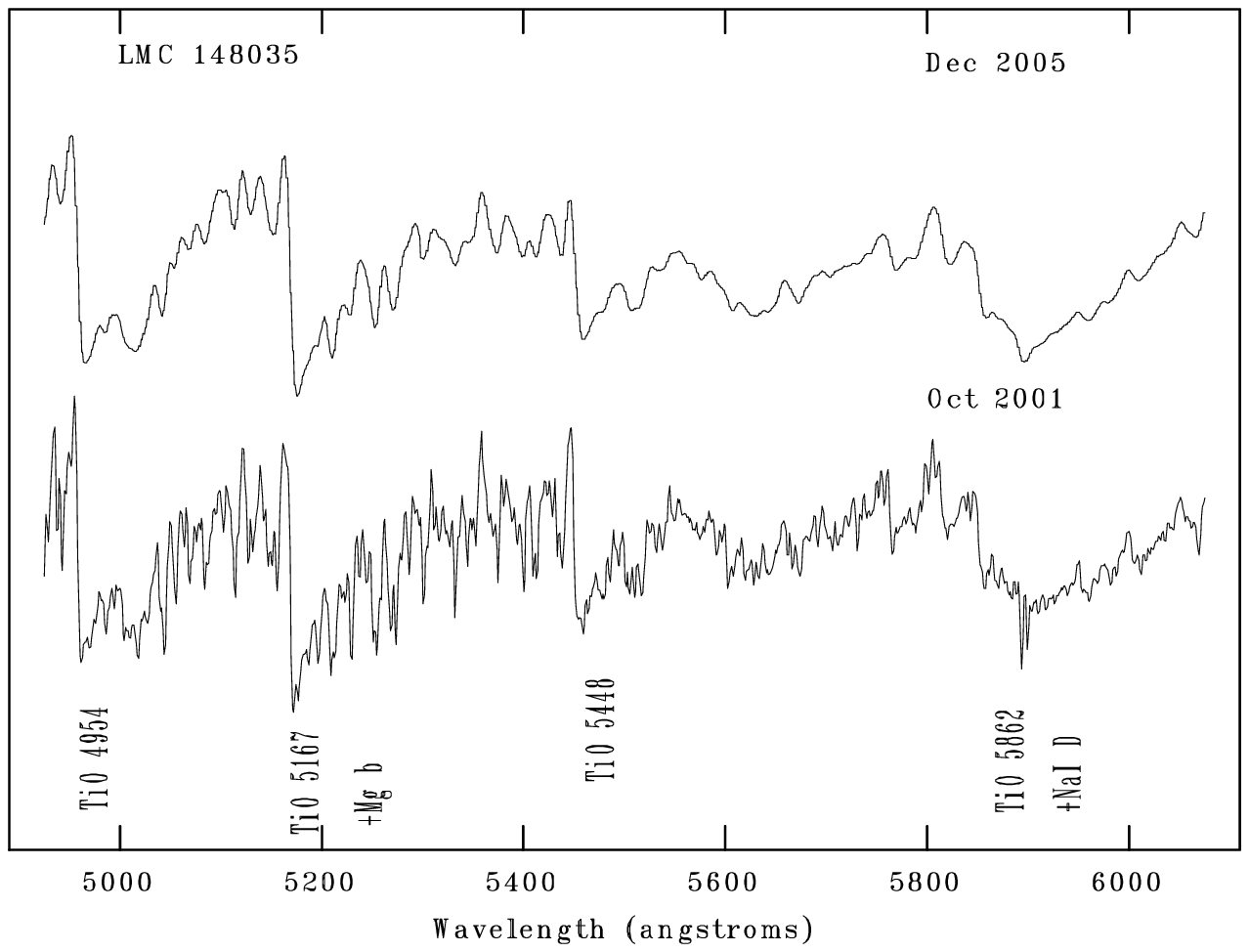}
\plotone{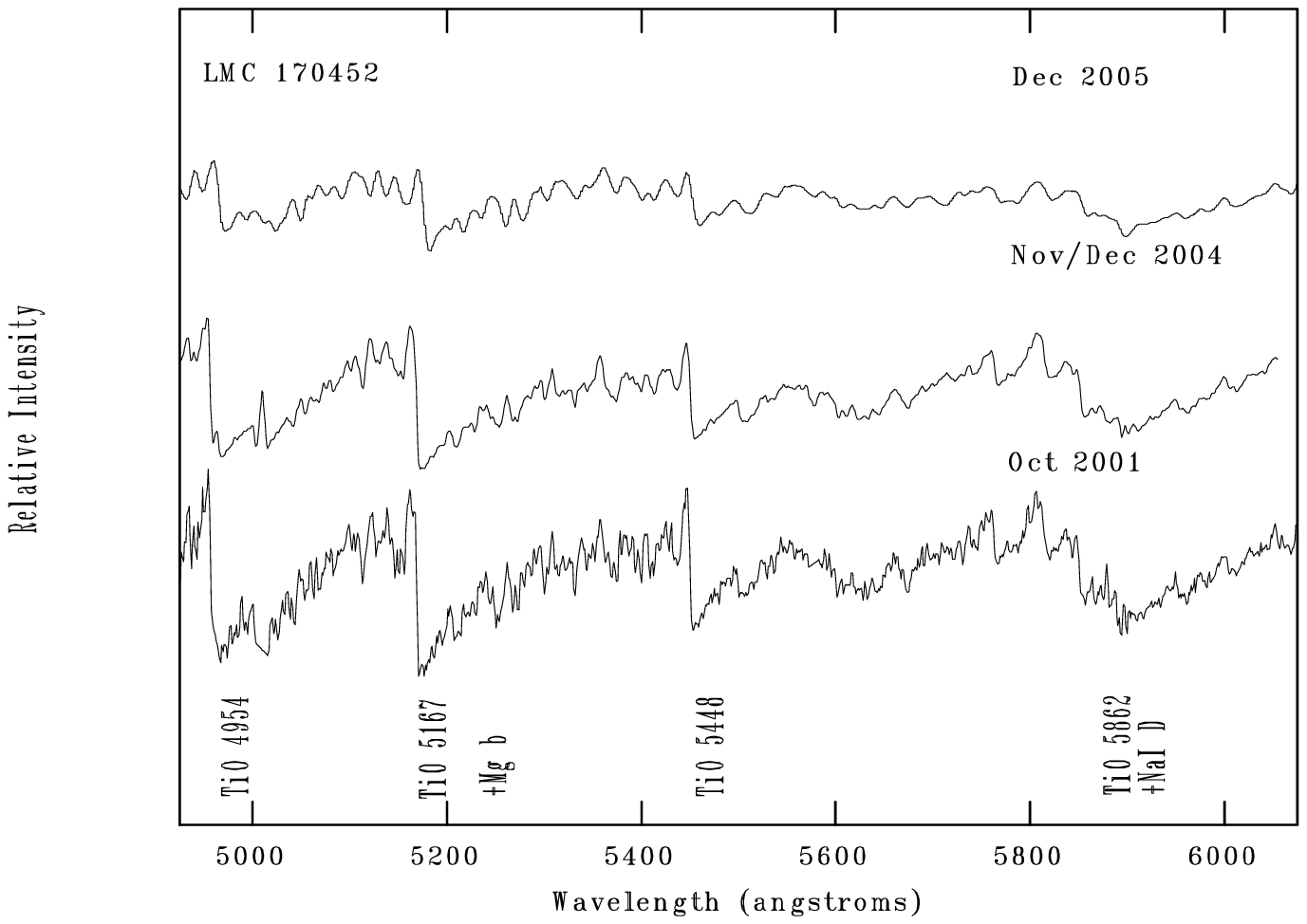}
\caption{\label{fig:changes} Spectral varaibility.  We show a comparison of the
October 2001 spectra (from Massey \& Olsen 2003) with those obtained
in December 2004 and December 2005 (from the present paper), where the spectra have been
normalized, and nebular emission has been removed from the October 2001
spectrum of LMC 170452 to make the comparison easier.
Prominent absorption features are
indicated, based upon the identifications given in Turnshek et al.\ (1985).
SMC 046662, SMC 055188, and LMC 170452 clearly show dramatic changes in the 
absorption line intensities, indicating a change in spectral type (and hence
effective temperature); the changes in LMC 148035 are relatively unremarkable.
}
\end{figure}

From examining the data in Table~\ref{tab:stars}, we find
four stars that show significant differences in their listed spectral types: SMC 046662 (M2~I to K2-3~I), SMC 055188 (M2~I to M4.5~I), LMC 148035 (M4~I to M2.5~I), and LMC 170452 (M4.5-5~I to M1.5~I).  We compare
their October 2001 spectra (from Massey \& Olsen 2003) with those from late
2004 and December 2005
 in Fig.~\ref{fig:changes}. We see that indeed the spectral
changes for SMC 046662, SMC 055188, and LMC 170452 are real, and that the line strengths have indeed changed dramatically throughout the past few
observations.  In the case of LMC 148035 the differences are much less
significant, and Massey \& Olsen (2003) have assigned too late a spectral type.

In the case of HV 11423 (Massey et al.\ 2007) we
were able to combine the changes in the photometric and  
spectrophotometric properties of the star
to present a picture of the physical changes that were taking place  
within the star. HV 11423 varied
in effective temperature from 4300 to 3500 K on a time-scale of  
months. When the star is as cool as it
gets, it has a very late spectral type, M4.5~I or so, much later than  
other supergiants that were known in the SMC (prior to this study),
and far beyond the region where the star is stable hydrodynamically.   
 This 800 K change in effective temperature was reflected in the
star's changing $V$ and luminosity: when the star was hot, it was
also brighter and slightly more luminous, with the differences
amounting to -0.6 mag in both cases. Although the
differences in $V$ and $M_{\rm bol}$ are the same, this is coincidental,
as the absolute {\it visual} magnitude $M_V$ changed by -1.9 mag. The
change in $V$ was smaller because our analysis also showed that the amount of
visual extinction also changed by 1.3 mag, due presumably to additional
circumstellar dust that forms when the star is cool. Of course, with only two or three epochs of  
observations it is difficult to sort out what
changes sporadically rather than systematically.

Still, a very similar picture seems to emerge here. For SMC 046662,  
SMC 055188, and LMC 170452 we see changes in
the spectral types and effective temperatures on the time scale of a  
year, albeit by lesser amounts. When these stars are hottest they are also
at their brightest. For SMC 046662 and  
SMC 055188 we can also conclude that when
the stars are hottest they are also more luminous (Table~\ref{tab:results}). The difference that we observed were smaller, with the changes in
effective temperature and bolometric luminosity amounting to
125 K and -0.2 mag, respectively, for SMC 046662. At the extremes we observed, the changes for SMC 055188 amounted to 200 K and -0.5 mag. Futhermore, we can estimate
the amount of extinction for these two stars, and find $A_V$ is larger when the star is  
hotter. All of this behavior is remarkably similar to what we see in HV 11423. 

\section{Results}
\label{Sec-results}

\subsection{Placement on the H-R Diagram}
\begin{figure}
\epsscale{1.2}
\plotone{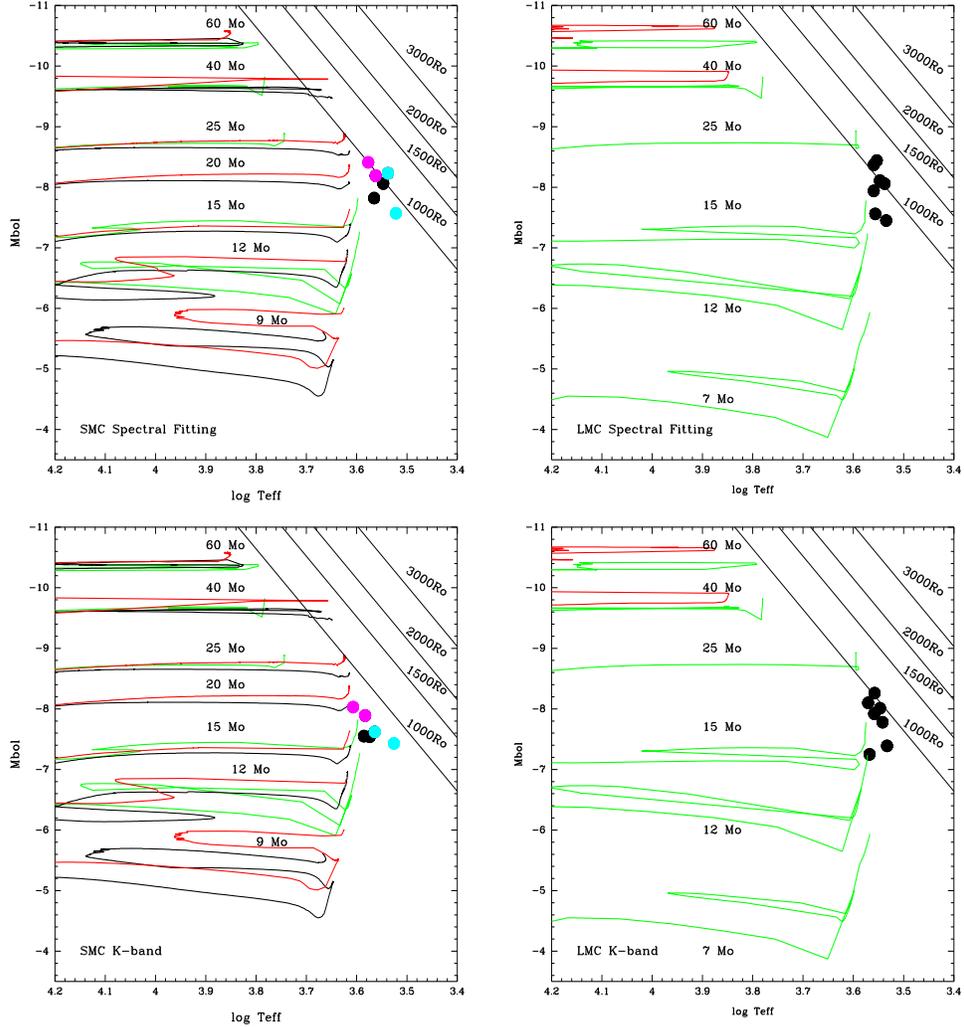}
\vskip -250pt
\caption{\label{fig:HRD} Location of the late-type MC RSGs discussed here compared to the evolutionary tracks.  We show the 
location of the RSGs in the H-R diagram of the SMC (left) and LMC (right), where the effective temperatures and bolometric
luminosities come from fitting the MARCS models to the optical spectrophotometry (top) and from the K-band photometry
(bottom).  The older, nonrotation evolutionary
tracks that include the effects of overshooting are shown in green and come
from Charbonnel et al.\ (1993) for the SMC, and from Schaerer et al.\ (1993) for the LMC.  The newer evolutionary
tracks (when available) are shown in black (zero rotation) and in
red (300 km s$^{-1}$ initial rotation) and come from Maeder \& Meynet (2001) for the SMC, and
Meynet \& Maeder (2005) for the LMC. Like-colored dots are used to link the
2004 and 2005 observations of the same star (purple for SMC 046662 and
light blue for SMC 055188).  
}
\end{figure}

In Figure~\ref{fig:HRD} we place our LMC and SMC sample stars on stellar evolutionary tracks of the
appropriate metallicity. It is clear that stellar evolutionary theory is not in agreement with
the observed parameters of these
late-type stars. Specifically, the evolutionary tracks do not extend to cool enough temperatures
to accomodate these stars in their current states. The location of the RSGs as derived from
spectral fitting is, on average, 275 K cooler than the tracks allow for the LMC and 541 K cooler
for the SMC.  The agreement is better for
the physical properties derived from $V-K$ photometry, although even here the
tracks do not extend to cool enough temperatures - the RSG locations are 205 K too cool for the LMC
and 216 K too cool for the SMC. However, this improved agreement is largely
due to the fact that, as discussed in (\S\ref{Sec-bb}), the temperatures derived from $V-K$
photometry are
also generally warmer than those derived from spectral fitting. The foreshadowing of this result
can be seen in Figure 8 of 
Paper II, where while agreement for most SMC and LMC RSGs is good, disagreement with
evolutionary theory is visible for the coolest SMC RSGs.   It appears, therefore,
that the location of these evolutionary tracks does not accomodate 
the full range of RSG properties in this low-metallicity environment.

The discrepancy appears slightly worse for the SMC than the LMC, particularly in the case of the
HRD positions derived from spectral fitting. Recall that
at low metallicity rotation plays an enhanced effect on the luminosities of the evolutionary
tracks (Maeder \& Meynet 2001) due to the effects of mixing.
Still, the current evolutionary models do not show much of a difference with the location in
temperature due to rotation, as is evident by comparing the black (no-rotation) and red 
(high rotation) tracks in Fig.~\ref{fig:HRD}. For the time being, this increased discrepancy seen
in the SMC remains unexplained.

\subsection{Revisions to the Effective Temperature Scale}
In Paper~II we compared the effective temperatures of stars of the same spectral subtype
in the SMC, LMC, and Milky Way.   Because the metallicity is less in the SMC than in the LMC
or Milky Way, we expect that a given band strength of TiO (the basis for the spectral classification)
will require a cooler temperature in the SMC than in the LMC or Milky Way, and indeed we
found such a progression: an M1 star would  have an effective temperature of 3625 K in the SMC,
3695 in the LMC, and 3745 in the Milky Way.  Put another way, stars with an effective temperature
of 3550 K would have TiO band strengths corresponding (roughly) to an M1.5~I in the SMC,
M2.5~I in the LMC, and M3.5~I in the Milky Way.  

\begin{figure}
\epsscale{0.8}
\plotone{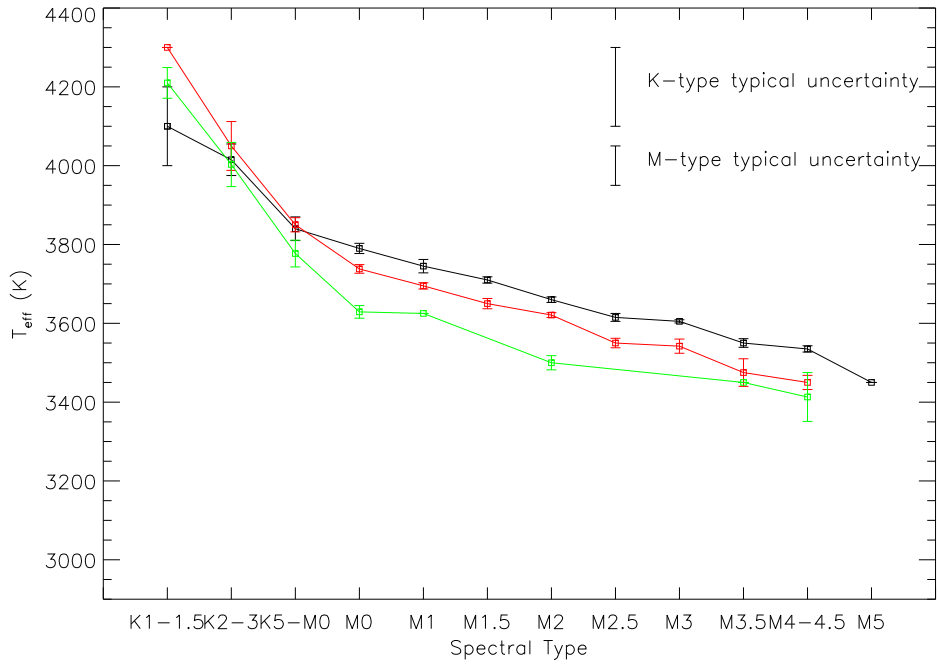}
\caption{\label{fig:TeffScale} Modified full temperature scales for the
Milky Way (black), LMC (red), and SMC (green), incorporating the eleven stars
from this sample. The
Magellanic Cloud scales now include more stars at the later spectral types,
which allows us to examine the metallicity effects at lower temperatures.
For example, a 3550 K star may appear as an M3.5 I in the Milky Way, an M2.5 I
in the LMC, and an M1.5 I in the SMC.}
\end{figure}

With the present data we can improve the comparison for the latest types.  In Table~\ref{tab:latetscale}
we update Table 4 of Paper~II. 
Fig.~\ref{fig:TeffScale} shows updated effective temperature scales for the LMC and SMC with these new late-type members
included. We have also included our results on HV 11423 (Massey et al.\ 2007). The error bars at the upper right show our estimate of the uncertainty when measuring the temperature of a single
star - 100~K for the early-K type stars and 50~K for the later-type RSGs, as described in \S\ref{Sec-spec}.  At M0-M2 we find that
a star in the LMC is about 50~K cooler than in the Milky Way, while a star in the SMC in
about 130~K cooler than in the Milky Way.

\section{Discussion}
\label{Sec-conclude}

We have determined the physical properties of a sample of seven late-type RSGs
in the LMC and 4 late-type RSGs in the SMC; one additional star with extreme properties, HV 11423,
 has been
studied separately (Massey et al.\ 2007).   We argue that these stars are likely all
members of the Magellanic Clouds and not foreground objects, and have found
that these stars possess photometric variability at $V$ that is larger than RSGs of
earlier spectral types. Although four of our stars show significant
variability in $B-V$, suggesting changes in the amount of circumstellar
reddening, the variable V-band magnitudes are not correlated with the color
changes and therefore must be due to physical changes in the star itself.
Consistent with this, three of these stars - SMC 046662, SMC 055188, and LMC 170452 - have demonstrated spectral
variability of several subtypes.  Other than the SMC star
HV 11423, this behavior is unknown for
RSGs. These late-type RSGs  are significantly cooler than the evolutionary tracks allow,
with the discrepancy larger for the SMC than the LMC.  Naively
we would argue that for the most part these stars are in the Hayashi
forbidden zone of the HRD, which is also true of HV 11423 when it is in its
cool state.

The extinction observed around most of these stars (Tables~\ref{tab:results} and~\ref{tab:phot}) is higher than what is typically seen for OB stars in the
Clouds, for which $A_V = 0.28$ (SMC), and $A_V = 0.40$ (LMC), where the values
come from Massey et al. (1995). This is similar to what was found by Massey
et al. (2005) for Galactic RSGs. Both HV~11423 and SMC 055188 are among the only four known
SMC RSGs that are IRAS sources (Massey et al.\ 2007), indicating that
we are seeing thermal emission from circumstellar dust.   In the case of
HV~11423, we found evidence of a variable amount of visual extinction, which
we argued was connected with the sporadic production of dust, and we now
find similar evidence for sporadic dust production when comparing the 2004
and 2005 results of model fitting for SMC 055188 and SMC 046662.
Like HV~11423, SMC 055188 probably produces dust quite sporadically:
despite its presence in the IRAS source catalog, the star was not in the Midcourse 
Space Experiment ({\it MSX}) 10$\mu$m flux  MSXC6 catalog (Egan et al.\ 2003)
despite the fact
that MSX was considerably more sensitive than IRAS at this wavelength.
Similar sporadic dust production may be true of other stars in this sample,
given that we find that four of our stars, (SMC 083593, LMC 148035, LMC 162635, and LMC
170452) show a change of several tenths of a magnitude in $B-V$ colors.
This could account for some of the $V$-band variability
we observe, but clearly not all, as in some cases $V$ has gotten larger while
$B-V$ has gotten smaller (e.g., SMC 083593, LMC 162635), or stayed the same
despite changes in $B-V$ (LMC 170452). Thus physical changes in the star are
primarily responsible for the variability in $V$, although changes in the
circumstellar extinction (as evidence by the $B-V$ changes) probably
complicated the light curves as well.

Most interestingly, HV~11423, originally thought to be a unique and
extreme case, has now been joined by three
fellow RSGs exhibiting similar behavior: cool stars, inhabiting the
Hayashi forbidden zone, that show large
variability in spectral type, $V$ magnitudes, and extinction, presumed to
be from circumstellar dust. These stars suffer changes in effective
temperature and bolometric luminosity on timescales of months; when
they are at their hottest, they are also brighter, dustier, and
more luminous. As
described above, one would expect stars in the Hayashi forbidden region to be
unstable hydrodynamically, which we expect to lead to this variability
and behavior. Further
monitoring of these stars, both photometrically and spectroscopically, may
lead to an improved understanding of this phase of massive star evolution.

\acknowledgements
We are grateful to CTIO for the hospitality and assistance provided during our observations. Constructive comments from the referee, Chris Evans, improved the
clarity and presentation of this manuscript. This 
paper made use of data from the Two Micron All Sky Survey, which is a joint project of the University
of Massachusetts and the Infrared Processes and Analysis Center/California Institute of Technology, 
funded by the National Aeronautics and Space Administration and the National Science Foundation. This
work was supported by the National Science Foundation through AST-0604569 to PM.

\clearpage

%these are currently the color figures; b&w figures are named f2abw.eps, etc.
\clearpage

\clearpage

\clearpage

\clearpage

\clearpage

\clearpage
\begin{deluxetable}{l c c c c c c c l l l l c }
\rotate
\tabletypesize{\scriptsize}
\tablewidth{0pc}
\tablenum{1}
\tablecolumns{13}
\tablecaption{\label{tab:stars} Program Stars}
\tablehead{
\colhead{Star}
& \colhead{$\alpha_{\rm 2000}$} 
& \colhead{$\delta_{\rm 2000}$}
& \multicolumn{4}{c}{Photometry\tablenotemark{a}}
& \colhead{}
& \multicolumn{4}{c}{Spectral Type}
& \colhead{Radial Velocity\tablenotemark{c}} \\ \cline{4-7} \cline{9-12} 
\colhead{[M2002]}
&\multicolumn{2}{c}{}
& \colhead{$V$}
& \colhead{$B-V$}
& \colhead{$K_S$}
& \colhead{$V-K$}
& \colhead{}
& \colhead{EFH85\tablenotemark{b}}
& \colhead{MO03\tablenotemark{c}}
& \colhead{New (2004)}
& \colhead{New (2005)}
& \colhead{(km s$^{-1}$)}
}
\startdata
SMC 046662 &00 59 35.04 &-72 04 06.2 &12.75  & 1.95 &8.35 &4.36 & &M0 Ia     &M2 I     &K2-3 I &M0 I & 180.2\\
SMC 052334 &01 01 54.16 &-71 52 18.8 &12.87  & 1.91 &8.69 &4.14 & &M0 Iab    &K7 I     &\nodata  &K5-M0 I & 165.5 \\
SMC 055188 &01 03 02.38 &-72 01 52.9 &14.23  & 2.22 &8.62 &5.57  &  &\nodata   &M2 I    &M4.5 I &M3-4 I & 176.8 \\
SMC 083593 &01 30 33.92 &-73 18 41.9 &12.83\tablenotemark{d} & 1.60 &8.60 &4.19 & &M2 Ia  &\nodata &\nodata &M2 I\tablenotemark{d} & 180.1 \\
LMC 143035 &05 29 03.58 &-69 06 46.3 &14.10  & 1.93 &7.90 &6.16 & &\nodata   &M3-4.5 I &M4.5-5 I &M4 I & 268.3\\
LMC 148035 &05 30 35.61 &-68 59 23.6 &13.42  & 1.98 &7.55 &5.83 & &\nodata   &M4 I     &\nodata &M2.5 I & 284.5\\
LMC 150040 &05 31 09.35 &-67 25 55.1 &13.09  & 1.90 &7.63 &5.42 & &\nodata    &M4 I     &M3-4 I &M3-4 I & 280.0\\
LMC 158646 &05 33 52.26 &-69 11 13.2 &13.80  & 2.18 &7.90 &5.86 & &\nodata   &M3-4 I   &\nodata &M2 I & 288.3 \\
LMC 162635 &05 35 24.61 &-69 04 03.2 &14.58  & 2.20 &8.66 &5.88 & &\nodata   &M1 I     &\nodata &M2 I & 296.9\\
LMC 168757 &05 37 36.96 &-69 29 23.5 &14.62  & 1.80 &8.21 &6.37 & &\nodata   &M3 I     &\nodata &M3-4 I & 272.3\\
LMC 170452 &05 38 16.10 &-69 10 10.9 &13.99  & 2.24 &7.87 &6.08 & &\nodata   &M4.5-5 I &M4.5-5 I &M1.5 I & 289.4 \\
\enddata
\tablenotetext{a}{$V$-band photometry is derived from the spectrophotometry
itself, while $K_S$ values are from 2MASS. The $B-V$ values are from 
Massey (2002).  The $V-K$ values have been
computed by transforming $K_S$ to the Bessell \& Brett 1988 ``standard" K
by using the Carpenter (2001) relationship $K=K_S+0.04$.}
\tablenotetext{b}{From Elias et al.\ 1985}
\tablenotetext{c}{From Massey \& Olsen 2003}
\tablenotetext{d}{This value was determined from the merged spectrum of our
2004 and 2005 spectra.}
\end{deluxetable}

\begin{deluxetable}{l c c c c c c c c c c c c c}
\rotate
\tabletypesize{\scriptsize}
\tablewidth{0pc}
\tablenum{2}
\tablecolumns{14}
\tablecaption{\label{tab:results} Physical Properties}
\tablehead{
\colhead{Star} &
\colhead{HJD - 2,450,000} &
\colhead{Type} &
\colhead{$V$} &
\colhead{$M_V$} &
\colhead{$A_V$} &
\multicolumn{4}{c}{Spectral Fitting} &
\colhead{} &
\multicolumn{3}{c}{$(V-K)_0$} \\ \cline{7-10} \cline{12-14}
\multicolumn{6}{c}{} &
\colhead{$T_{\rm eff}$} &
\colhead{log $g$ (Model)} &
\colhead{$R/R_\odot$} &
\colhead{$M_{\rm bol}$} &
\colhead{} &
\colhead{$T_{\rm eff}$} &
\colhead{$R/R_\odot$} &
\colhead{$M_{\rm bol}$}
}
\startdata
SMC 046662 &3340.58 &K2-3 I  &12.38  &-7.14  &0.62 &3775 &0.0   &1000 &-8.41 & &4050 &730  &-8.03  \\
           &3725.63 &M0 I    &12.75  &-6.65  &0.50 &3650 &0.0   &960  &-8.19 & &3829 &760  &-7.89  \\
SMC 052334 &3725.67 &K5-M0 I &12.87  &-6.34  &0.31 &3675 &0.0   &800  &-7.82 & &3851 &640  &-7.55  \\
SMC 055188 &3341.59 &M4.5 I    &14.99  &-4.87  &0.96 &3325 &0.0   &870  &-7.57 & &3355 &800  &-7.43  \\               
           &3725.68 &M3-4 I  &14.23  &-6.07  &1.40 &3450 &0.0   &1100 &-8.23 & &3667 &730  &-7.62  \\
SMC 083593 &\nodata\tablenotemark{a} &M2 I    &12.83  &-6.16  &0.09 &3525 &0.0   &970  &-8.06 & &3751 &680  &-7.54  \\
LMC 143035 &3725.81 &M4 I    &14.10  &-5.64  &1.24 &3450 &-0.5  &1020 &-8.06 & &3482 &870  &-7.78  \\
LMC 148035 &3725.81 &M2.5 I  &13.42  &-6.57  &1.49 &3575 &0.0   &1130 &-8.44 & &3612 &1010 &-8.26  \\
LMC 150040 &3725.82 &M3-4 I  &13.09  &-6.03  &0.62 &3525 &-0.5  &990  &-8.11 & &3523 &950  &-8.01  \\
LMC 158646 &3725.82 &M2 I    &13.80  &-6.25  &1.55 &3625 &-0.5  &870  &-7.94 & &3619 &870  &-7.92  \\
LMC 162635 &3725.83 &M2 I    &14.58  &-5.78  &1.86 &3600 &0.0   &740  &-7.56 & &3696 &610  &-7.25  \\
LMC 168757 &3727.84 &M3-4 I  &14.62  &-4.90  &1.02 &3425 &0.0   &780  &-7.45 & &3411 &760  &-7.39  \\
LMC 170452 &3727.83 &M1.5 I  &13.99  &-6.68  &2.17 &3625 &0.0   &1060 &-8.37 & &3720 &890  &-8.10  \\
\enddata
\tablenotetext{a}{From merged 2004 and 2005 data.}
\end{deluxetable}

\begin{deluxetable}{l c c c c c}
\tablewidth{0pc}
\tabletypesize{\scriptsize}
\tablenum{3}
\tablecolumns{6}
\tablecaption{\label{tab:phot} Time Resolved Data\tablenotemark{a}}
\tablehead{
\colhead{HJD-2,450,000}
&\colhead{$V$\tablenotemark{b}}
&\colhead{$A_V$}
&\colhead{$B-V$}
&\colhead{$T_{\rm eff}$}
&\colhead{Spectral Type}
}
\startdata
\cutinhead{[M2002] SMC 046662}
1186.58 & 12.90 &\nodata & 1.88 &\nodata &\nodata \\
2188.56 & 13.32 &\nodata &\nodata &\nodata &M2 I \\ %2186
3340.58 & 12.38 &0.62 & \nodata &3775 & K2-3 I\\
3725.63 & 12.75 &0.50 & 1.95 &3650 & M0 I\\
\cutinhead{[M2002] SMC 052334}
1186.58 &   12.89 &\nodata & 1.94 &\nodata &\nodata \\
2187.56 &   12.72 &\nodata &\nodata &\nodata &K7 I \\ %2187
3725.67 &   12.87 &0.31 & 1.91 &3675 &K5-M0 I\\
\cutinhead{[M2002] SMC 055188}
1186.58  & 14.96 &\nodata & 2.25 &\nodata &\nodata\\
2188.56  &\nodata &\nodata &\nodata &\nodata &M2 I \\ %nodata
3341.59  & 14.99 &0.96 & \nodata &3325 &M4.5 I\\ 
3725.68 &  14.23 &1.40 & 2.22 &3450 &M3-4 I\\
\cutinhead{[M2002] SMC 083593}
1186.59  & 12.64 &\nodata & 1.87 &\nodata &\nodata\\
3334.60  & 12.81 &\nodata & \nodata &\nodata &\nodata\\ 
3724.60\tablenotemark{c}  & 12.83 &0.09 & 1.60 &3525 &M2 I\\
\cutinhead{[M2002] LMC 143035}
1186.67  &  13.53 &\nodata & 1.93 &\nodata &\nodata\\
1998.51  & 13.51 &\nodata & 1.93 &\nodata &\nodata\\
2188.72  & 14.24 &\nodata &\nodata &\nodata &M3-4.5 I \\ %2184
3334.60   & 13.99 &\nodata & \nodata &\nodata &M4.5-5 I\\
3725.81  &  14.10 &1.24 & 1.93 &3450 &M4 I\\
\cutinhead{[M2002] LMC 148035}
1186.67   & 12.94 &\nodata & 1.82 &\nodata &\nodata\\
1996.53   & 13.89 &\nodata & 1.71 &\nodata &\nodata \\
1998.51   & 13.87 &\nodata & 1.62 &\nodata &\nodata \\
2188.72   & 12.94 &\nodata &\nodata &\nodata &M4 I \\ %2187
3725.81    & 13.42 &1.49 & 1.98 &3575 &M2.5 I\\
\cutinhead{[M2002] LMC 150040}
1186.71  & 12.86 &\nodata & 1.97 &\nodata &\nodata\\
1997.54   & 12.81 &\nodata & 1.96  &\nodata &\nodata\\
2188.82   & 12.93 &\nodata &\nodata &\nodata &M4 I \\ %2187
3334.60    & 12.60 &\nodata & \nodata &\nodata &M3-4 I\\ 
3725.82   & 13.09 &0.62 & 1.90 &3525 &M3-4 I\\
\cutinhead{[M2002] LMC 158646}
 1186.71   & 12.66 &\nodata & 2.19 &\nodata &\nodata\\
1996.53    & 13.10 &\nodata & 2.23 &\nodata &\nodata\\
2188.79    & 13.54 &\nodata &\nodata &\nodata &M3-4 I \\ %2187
 3725.82    & 13.80 &1.55 & 2.18 &3625 &M2 I\\
\cutinhead{[M2002] LMC 162635}
1996.53   & 14.23 &\nodata & 2.33 &\nodata &\nodata\\
2188.79   & 13.52 &\nodata &\nodata &\nodata &M1 I \\ %2186
3725.83   & 14.58 &1.86 & 2.20 &3600 &M2 I \\
\cutinhead{[M2002] LMC 168757}
1996.53  &  14.08 &\nodata & 1.77 &\nodata &\nodata\\
2188.76   & 13.23 &\nodata &\nodata &\nodata &M3 I \\ %2186
3727.84   &  14.62 &1.02 & 1.80 &3425 &M3-4 I \\
\cutinhead{[M2002] LMC 170452}
1996.53   &  13.99 &\nodata & 2.39 &\nodata &\nodata\\
2188.76   &\nodata &\nodata &\nodata &\nodata &M4.5-5 I \\ %nodata
3341.70   &  15.31 &\nodata & \nodata &\nodata &M4.5-5 I \\
3727.83    & 13.99 &2.17 & 2.24 &3625 &M1.5 I \\
\enddata
\tablenotetext{a}{Measurements from 2,451,186-2,451,999 were obtained
using the CCD images described in Massey (2002).  The uncertainty in
these values are 0.01~mag or smaller. Later measurements come
from our 2004 November/December and 2005 December spectrophotometry.}
\tablenotetext{b}{The $V$ magnitudes quoted for the 2,452,188 HJD's
comes from ASAS observations on the closest approximate date
to the observations, from 2,452,184-2,452,187.}
\tablenotetext{c}{From merged 2004 and 2005 data.}
\end{deluxetable}

%\begin{deluxetable} {l c c c}
%\tabletypesize{\footnotesize}
%\tablewidth{0pc}
%\tablenum{4}
%\tablecolumns{4}
%\tablecaption{\label{tab:deltas} Title?}
%\tablehead{
%\colhead{HJD} &
%\colhead{$V$} &
%\colhead{$(B-V)$} &
%\colhead{Spectral Type} \\
%}
%\startdata
%\cutinhead{SMC 046662}
%1187 & 12.90 &1.88 &\nodata \\
%2187 & ASAS  &\nodata &M0~I \\
%3341 & 12.38 &\nodata &K5-M0~I \\
%3725 & 12.75 &1.95 &M2~I \\
%\cutinhead{LMC 170452}
%1997 & 13.99 &2.38 &\nodata \\
%2187 & ASAS &\nodata &M4.5-5~I \\
%3342 & 15.31 &\nodata &M4.5-5~I \\
%3728 & 13.99 &2.24 &M1.5~I \\
%\enddata
%\end{deluxetable}

\begin{deluxetable} {l c c c c c c c c c c c}
\tabletypesize{\footnotesize}
\tablewidth{0pc}
\tablenum{4}
\tablecolumns{12}
\tablecaption{\label{tab:latetscale}Effective Temperature Scales with New Data}
\tablehead{
\colhead{} &
\multicolumn{3}{c}{SMC} & &
\multicolumn{3}{c}{LMC} &&
\multicolumn{3}{c}{Milky Way\tablenotemark{a}} \\ \cline{2-4} \cline{6-8} \cline{10-12}
\colhead{Spectral Type} & 
\colhead{$T_{\rm eff}$ (K)}  &  
\colhead{$\sigma_\mu$\tablenotemark{b}} &
\colhead{$N$} & &
\colhead{$T_{\rm eff}$ (K)} &
\colhead{$\sigma_\mu$\tablenotemark{b}} &
\colhead{$N$} & &
\colhead{$T_{\rm eff}$ (K)} &
\colhead{$\sigma_\mu$\tablenotemark{b}} &
\colhead{$N$} \\
}
\startdata
K1-K1.5 I & 4210    & 39      & 7        && 4300    & \nodata & 1       && 4100 & 100     & 3 \\
K2-K3 I   & 4003    & 56      & 15       && 4050    & 62      & 3       && 4015 & 40      & 7 \\
K5-M0 I   & 3777    & 34      & 11       && 3850    & 18      & 2       && 3840 & 30      & 3 \\
M0 I      & 3629    & 16      & 6        && 3738    & 11      & 4       && 3790 & 13      & 4 \\
M1 I      & 3625    & \nodata & 1        && 3695    & 8       & 5       && 3745 & 17      & 7 \\
M1.5 I    & \nodata & \nodata & \nodata  && 3650    & 13      & 7       && 3710 & 8       & 6  \\
M2 I      & 3500    & 18 & 2        && 3621    & 6       & 7       && 3660 & 7       & 17  \\
M2.5 I    & \nodata & \nodata & \nodata  && 3550    & 12      & 6       && 3615 & 10      & 5 \\
M3  I     & \nodata & \nodata & \nodata  && 3542    & 18      & 3       && 3605 & 4       & 9  \\
M3.5 I    & 3450    & \nodata & 1        && 3475    & 35      & 2       && 3550 & 11      & 6  \\
M4-M4.5 I & 3413    & 62      & 2        && 3450    & 18      & 3       && 3535 & 8       & 6  \\
M5 I      & \nodata & \nodata & \nodata  && \nodata & \nodata & \nodata && 3450 & \nodata & 1  \\

\enddata
\tablenotetext{a}{From Paper I.}
\tablenotetext{b}{Standard deviation of the mean.}
\end{deluxetable}

\end{document}